\documentclass[12pt,onecolumn]{IEEEtran}
\usepackage[top=0.95in, bottom=0.95in, left=0.95in, right=0.95in]{geometry} % 控制页面边距，是设置行宽的关键
\usepackage{setspace}
\usepackage[usenames]{color}

\usepackage{cite,amsmath,amsfonts,amssymb,amsthm,mathrsfs}
\usepackage{bm}
\usepackage{epsfig,graphicx,subfigure}
\theoremstyle{remark}

\usepackage{mathtools}

\usepackage[colorinlistoftodos]{todonotes}
\usepackage{cleveref}
\usepackage[ruled]{algorithm2e}
\usepackage{algorithmic}
\usepackage{stfloats}
\usepackage[justification=centering]{caption}
\usepackage{booktabs}
\usepackage{multirow}
\usepackage{extarrows}
\usepackage{makecell}
\usepackage{threeparttable}

\begin{document}
\title{Embodied AI-Enhanced IoMT Edge Computing: UAV Trajectory Optimization and Task Offloading with Mobility Prediction}
\author{
Siqi Mu, Shuo Wen, Yang Lu,~\IEEEmembership{Member,~IEEE}, Ruihong Jiang,~\IEEEmembership{Member,~IEEE}, and Bo Ai,~\IEEEmembership{Fellow,~IEEE} % \normalsize

\thanks{S. Mu and S. Wen are with the School of Sports Engineering,
Beijing Sport University, Beijing 100084, China
(e-mail: musiqi@bsu.edu.cn).}
\thanks{Y. Lu is with the School of Computer and Information Technology, Beijing Jiaotong University, Beijing 100044, China (e-mail: yanglu@bjtu.edu.cn).}
\thanks{R. Jiang is with the State Key Laboratory of Networking and Switching Technology, Beijing University of Posts and  Telecommunications, Beijing 100876, China (e-mail: rhjiang@bupt.edu.cn).}
\thanks{B. Ai are with the School of Electronics and Information Engineering, Beijing Jiaotong University, Beijing 100044, China (e-mail: boai@bjtu.edu.cn).}
\thanks{Corresponding author: Yang Lu.}
}

% \graphicspath{{Figs/}}
% \pagestyle{empty}

\maketitle

\begin{abstract}
  Due to their inherent flexibility and autonomous operation, unmanned aerial vehicles (UAVs) have been widely used in Internet of Medical Things (IoMT) to provide real-time biomedical edge computing service for wireless body area network (WBAN) users. In this paper, considering the time-varying task criticality characteristics of diverse WBAN users and the dual mobility between WBAN users and UAV, we investigate the dynamic task offloading and UAV flight trajectory optimization problem to minimize  the weighted average task completion time of all the WBAN users, under the constraint of UAV energy consumption. To tackle the problem, an embodied AI-enhanced IoMT edge computing framework is established. Specifically, we propose a novel hierarchical multi-scale Transformer-based user trajectory prediction model based on the users' historical trajectory traces captured by the  embodied AI agent (i.e., UAV). Afterwards, a prediction-enhanced deep reinforcement learning (DRL) algorithm that integrates predicted users' mobility information is designed for intelligently optimizing  UAV flight trajectory and task offloading decisions. Real-word movement traces and simulation results demonstrate the superiority of the proposed methods in comparison with the existing benchmarks.

\end{abstract}
%*********************************************************************
\section{Introduction}\label{sec:introduction}
%*********************************************************************
%*********************************************************************
\subsection{Background and Prior Works}\label{sec:background}
%*********************************************************************
Recent advancements in Internet of Medical Things (IoMT) and artificial intelligence (AI), have made a significant contribution to sustainable digital health. Combining traditional medical equipments with IoT, IoMT provides ubiquitous in-home healthcare, and greatly alleviates public medical burdens and saves healthcare resources\cite{philip2021internet}. As the key components of IoMT, wireless body area networks (WBANs) deploy various low-power biosensors on numerous people. These heterogeneous biosensors sense various types of physiological data, including electrocardiogram
(ECG), electroencephalogram (EEG), blood pressure (BP), body temperature etc., and transmit
the data to an on-body sink node for further processing\cite{movassaghi2014wireless, mu2025aoi}. WBANs have significantly facilitated pervasive health monitoring services and promoted real-time
health assessment.

Despite these promising developments, the rapid population growth, especially the aged, and their medical tasks still
overloads the healthcare infrastructure, and limits the development of
IoMT\cite{ning2020mobile}. Local sink nodes, such as mobile phones and laptops,
cannot satisfy the latency requirements of the time-sensitive tasks
for medical information analysis. The proliferation of mobile
edge computing (MEC) is conceived as a promising paradigm
for tackling such challenges\cite{lu2025agentic}. By providing computation resources for the offloaded medical analysis
tasks in proximity, MEC alleviates the burden on local devices
and augments the capability of IoMT. The integration of IoMT and MEC has been envisioned as an effective approach for real-time healthcare service provision, especially during the COVID-19 pandemic\cite{zhu2022iomt,rahman2021internet}.

% Considering the mobility of WBAN users,
With the continuous increase in WBAN users, cellular infrastructure-based MEC, in which edge servers deployed at terrestrial base station, struggles to provide seamless connectivity and reliable computation\cite{mao2024uav}. Benefitting from its high mobility, flexible
deployment capabilities and strong scalability, UAV-enabled MEC has
gained widespread attention and become a research hotspot.
% In spite of its prominent advantages, UAV-enabled MEC faces practical challenges, such as finite battery life and limited computation resources. Considering these constraints,
Particularly,
computation offloading and UAV flight trajectory
in a MEC system are jointly optimized to minimize the overall task delay of users\cite{hu2018joint,guo2019joint}, or minimize the energy consumption of UAV\cite{sun2021joint}. By enabling UAV to simultaneously act as a relay and MEC server, \cite{liu2023maximizing} and \cite{hu2019uav} optimized the task offloading, bandwidth allocation, computation resource scheduling and UAV trajectory using successive convex approximation, such that maximizing the energy efficiency of users and UAV. Later in \cite{bai2022delay},  a UAV enabled edge-cloud computing system was considered to augment the computation capability of the UAV. A delay minimization problem for edge-cloud cooperative offloading was investigated in this paper.
However, these earlier studies have been conducted on  the static scenarios without consideration on user mobility, which is unrealistic since user locations may change dynamically over time in practice. In addition, user mobility directly impacts UAV path planning and edge computing performance.

To tackle this problem, several studies focused on mobile users, where the user mobility model follows the random waypoint mobility model\cite{amer2020mobility}, reference point group mobility model\cite{wang2023joint,yan2023joint}, or the Gauss-Markov mobility model\cite{omoniwa2022optimizing,yang2021dynamic}. These models have been applied to derive the UAV coverage probability\cite{amer2020mobility} or develop optimization algorithms on user association, power allocation, subchannel assignment, UAV positioning or trajectory design under various system objectives, such as maximizing system throughput\cite{yan2023joint,wang2023joint} or energy efficiency\cite{omoniwa2022optimizing, yang2021dynamic}.
However, in these ideal mobility models,
the direction of user's movement tends to be uniformly
distributed among left, right, forward and backward, which
does not fully reflect the complexities and nuances  of the real-world user movement \cite{liu2019trajectory}. Additionally,
although user mobility patterns were considered, the algorithm designs in these works were based on the assumption that precise user locations are accessed by UAV in real time. Such an assumption is difficult to fulfill in practice, especially in urban environments or areas with significant obstructions. Even worse, the user location information reported to UAV may be outdated due to the fast movement of users, leading to suboptimal task offloading strategy and UAV path planning\cite{wu2024deep}.

In view of these, a few researchers have made efforts to capture the time-varying uncertainty of user mobility with prediction models to improve the service quality of edge computing. \cite{ma2020leveraging} proposed a LSTM-based  mobility prediction model, based on which a predictive service placement algorithm was designed to balance the latency performance and handover cost. In \cite{wu2023mobility}, the authors developed a seq2seq user trajectory prediction model,
 alongside a deep reinforcement learning (DRL) algorithm for supporting offloading decisions and resource allocation in MEC, to minimize the average task latency of users.
In spite of these innovative attempts, several challenges still remains. First, these existing studies on MEC with mobility prediction
mainly focus on the communication connections between terrestrial base station and users, how
to improve the edge computing performance in an air-ground system with dual mobility of UAV and ground users needs to be further investigated. Besides, inaccuracies in the existing trajectory prediction methods can result in suboptimal offloading decisions and UAV trajectory optimization, which will increase  task completion time and UAV energy consumption. It is imperative to develop a more robust and predictive framework.
Finally, few works on UAV-assisted MEC systems
consider the time-varying criticality of computation tasks, which is a significant and indispensable feature for WBAN users. How to intelligently make offloading decisions based on the time-varying task characteristics is non-trivial.
%

%*********************************************************************
\subsection{Contributions}\label{sec:contribution}
%*********************************************************************
Motivated by the challenges and inspired by the advanced AI techniques, we propose an  embodied AI-enhanced UAV edge computing framework in this work. Embodied AI, which emphasizes the physical objects embedded with intelligent system actively interact with and  learn from their physical surroundings, has been shown a promising solution for dealing with this highly complex and dynamic scenario\cite{zhang2025embodied, zhang2024generative}.
Specifically, the proposed framework is comprised of two core modules, i.e., a hierarchical Transformer enabled user mobility prediction module and a DRL enabled UAV trajectory optimization and task offloading module. The embodied AI system embedded within UAV enables accurate mobility prediction and real-time strategy adaptation to dynamic environments. Equipped with the designed AI algorithms, the UAV embodied AI agent predicts user mobility
based on the perceived historical information, and autonomously optimizes flight trajectory and makes intelligent task offloading decisions.  The contributions of this work are summarized as follows:

\begin{itemize}
  \item[1)] Considering the time-varying task criticality characteristics of diverse WBAN users, we formulate a dynamic multi-stage task offloading and UAV flight trajectory optimization problem, aiming at minimizing the weighted average task completion time of all the WBAN users, subject to the total energy consumption of the UAV.
  \item[2)]To facilitate the optimization of flight trajectory and task offloading decisions for the UAV embodied AI agent, we propose a novel user trajectory prediction model based on a hierarchical multi-scale Transformer framework. Through the design of trajectory slice partitioning, embedding representation and the attention mechanism, the proposed model can capture the temporal dependencies of historical user trajectory on various time scales.
  \item[3)] The original dynamic optimization problem is transformed into a Markov decision process (MDP) problem. Based on the designed state, action and reward function, a prediction-enhanced DRL algorithm that integrates predicted users' mobility information is developed for intelligent UAV trajectory optimization and task offloading.
  \item[4)] We evaluate the performance of the proposed mobility prediction model and DRL  algorithm. Real-word traces  and  simulation results demonstrate that the proposed methods are superior in both effective mobility prediction and  optimizations on task offloading and UAV flight trajectory compared with the existing benchmarks.
\end{itemize}

The organization of this paper is as follows.
In Section \ref{sec:model}, the system model is introduced and the multi-stage
optimization problem is formulated.  Section \ref{sec:mobility} presents the hierarchical multi-scale Transformer framework for mobility prediction. Section \ref{sec:DRL} provides the prediction-enhanced UAV trajectory optimization and task offloading algorithm.  Performance evaluation
results are shown in Section \ref{sec:eva}. Finally, Section \ref{sec:con} concludes our work and points out possible future work.
% A summary of the primary
% symbols and notations employed in this paper are provided in TABLE \ref{table.notations}.

%*********************************************************************
\section{System Model}\label{sec:model}
%*********************************************************************
%
\begin{figure}[t]
\centering
\includegraphics[width=4in]{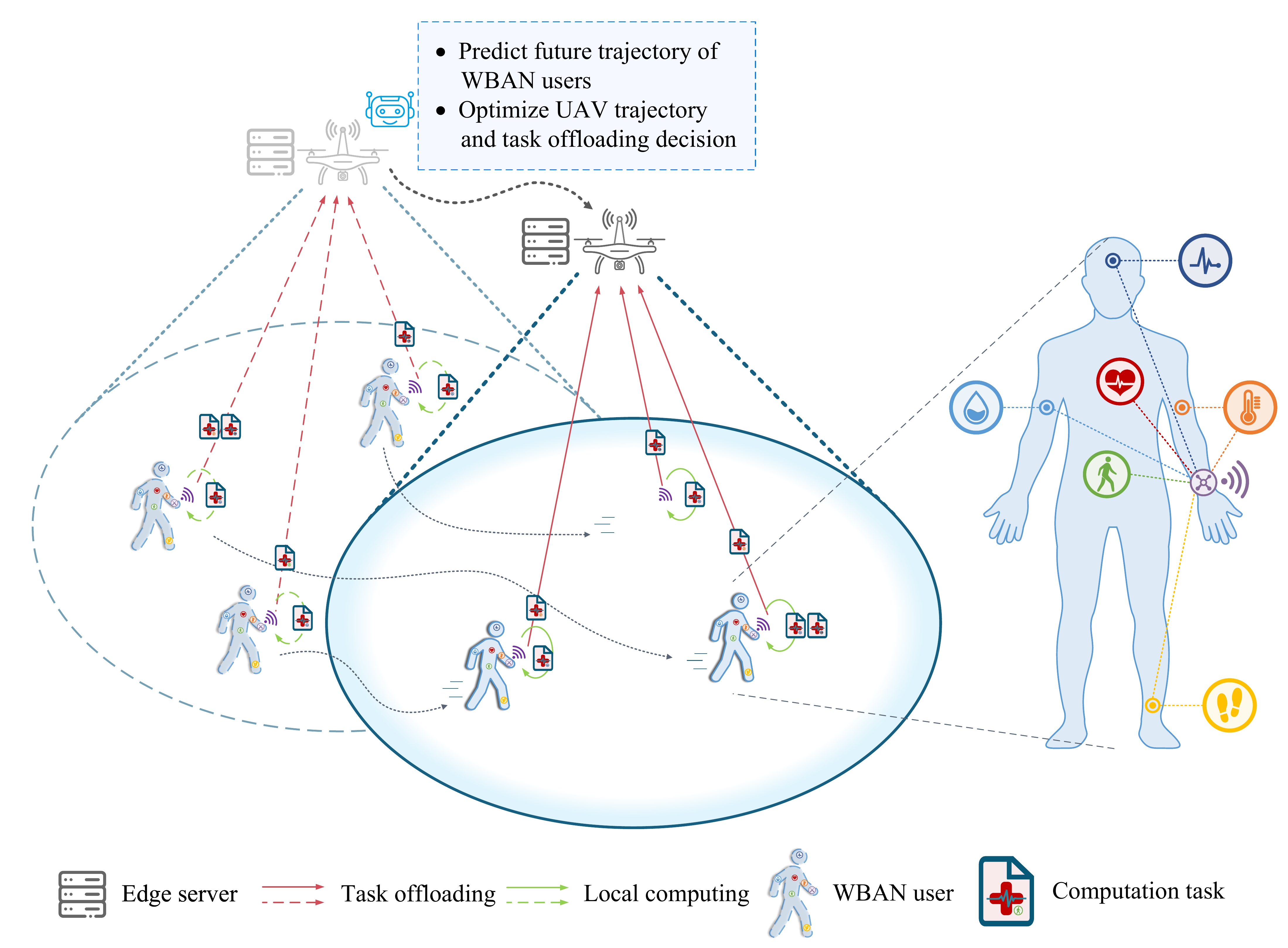}
\caption{System model}
\label{fig:model}
\end{figure}
We consider a UAV-enabled WBAN edge computing system as illustrated in Fig. \ref{fig:model}, where a UAV equipped with an edge server has a mission to provides the proximate computation service for mobile WBAN users. The set of WBAN users is denoted as $\mathcal{U}=\{1,\ldots, U\}$ and each user has $N$ heterogeneous computation tasks from its equipped biosensors, indexed as $n\in \mathcal{N}=\{1,\ldots,N\}$. The mission period of UAV is denoted as $T^{\max}$, which is slotted into $T$ time slots, represented as $t\in \mathcal{T}= \{1,\ldots,T\}$, with time slot duration of $\tau$.
 Considering a 3-D Cartesian coordinate system, user $u\in\mathcal{U}$ has a zero altitude and its horizontal location at time slot $t$ is denoted as $\bm{p}_u[t]=(x_u[t],y_u[t])$. We assume the UAV flies at a fixed altitude $H$, and the initial  horizontal locations of the UAV is preset as $\bm{p}_v[1]=(x_I,y_I)$.
 At the beginning of each time slot, WBAN users generate tasks and the UAV moves to the next location based on the observed locations and predicted user mobility.  After relocating a new location of the UAV, WBAN users offload a portion of tasks via wireless links, and the UAV assists the execution of these computation tasks.
 At time slot $t$, the horizontal location of the UAV is denoted as $\bm{p}_v[t]=(x[t],y[t])$. It is assumed that the UAV flies with a constant speed $v[t]$ at time $t$ and the direction of flight
is represented by $\sigma[t]$. Then, the UAV flying from
the previous hover location to the new location can be
expressed as
\begin{align}
  \bm{p}_v[t+1]
  =[x[t]+v[t]t^{\text{fly}}\cos\sigma[t], y[t]+v[t]t^{\text{fly}}\sin\sigma[t]],
\end{align}
where $t^{\text{fly}}$ is the UAV flying time in each time slot.
%*********************************************************************
\subsection{Computation Task Model}\label{sec:task}
Due to the varying sensed physiological states by each biosensor, the tasks of WBAN users are dynamically changed across time slots. For WBAN user $u$, its computation task $n$ at time slot $t$ is represented as a tuple $\Theta_{u,n}[t]=\langle{I_{u,n}[t],D_{u,n}[t],C_{u,n}[t]\rangle}$, where $I_{u,n}[t]$ is the current criticality index of task $n$, $D_{u,n}[t]$ is the data load of task $n$, and $C_{u,n}[t]$ the computation amount of task $n$.

Specifically, criticality index $I_{u,n}[t]$ is comprised of three parts, $\phi_u$, $\rho_{u,n}$ and $\alpha_{u,n}[t]$, where $\phi_u$ and $\rho_{u,n}$ are introduced to model the criticality of WBAN user $u$ and its biosensor $n$, and $\alpha_{u,n}[t]$ represents the importance of the sensing data of biosensor $n$\cite{askari2021energy}. The larger the values of $\phi_u$ and $\rho_{u,n}$, the higher the data criticality of the corresponding user and its biosensor. For example, the data for heart disease patients have a higher criticality than healthy users, and the value of $\rho_{u,n}$ for ECG biosensor used to monitor heart is greater than that of EMG biosensor used to  monitor muscle activity.
% In the system deployment stage, $\phi_u$ and $\rho_{u,n}$ of each user and its biosensor can be normalized to fixed values in $[0,1]$ according to the number of user categories and biosensor categories.
%
Besides, the data of the same biosensor can be divided into normal data and emergency abnormal data. The data importance of biosensor  $\alpha_{u,n}[t]$ indicates that the urgency of the sensing typical data $\theta_{u,n}[t]$. The  predefined normal value range is $[\breve{\theta}_{u,n}, \hat{\theta}_{u,n}]$.
 % and define the difference between the upper bound and the lower bound of the normal value as $\Gamma_{u,n}=\hat{\theta}_{u,n}-\breve{\theta}_{u,n}$.
Without loss of generality, $\alpha_{u,n}[t]$ is classified as two levels, i.e., low and high. If $\theta_{u,n}[t]$ is within the predefined normal value range $[\breve{\theta}_{u,n}, \hat{\theta}_{u,n}]$, it indicates the low urgency. Otherwise, it represents an abnormal states with high urgency.
 Thus, $\alpha_{u,n}[t]$ can be expressed as
 \begin{align}
\alpha_{u,n}[t]
=\left\{ \begin{array}{l} \mbox{low},  \mbox{if} \ \theta_{u,n}[t]\in [\breve{\theta}_{u,n}, \hat{\theta}_{u,n}], \\[6pt]
 \mbox{high},  \mbox{if} \ \theta_{u,n}[t]\in (-\infty,\breve{\theta}_{u,n}) \cup (\hat{\theta}_{u,n},+\infty).
     \end{array} \right.
 \end{align}

By jointly considering the user categories, biosensor categories and data importance, criticality index $I_{u,n}[t]$ of biosensor $n$ for WBAN user $u$ is defined as a function of the three factors, written as  $I_{u,n}[t]=\mathcal{F}(\phi_u,\rho_{u,n},\alpha_{u,n}[t])$.

Each task is atomically indivisible and can be processed locally or offloaded to the UAV for computing. Let $z_{u,n}[t]\in\{0,1\}$ denote as the indicator of the task offloading decision for task $\Theta_{u,n}[t]$. $z_{u,n}[t]=0$ signifies that task  $\Theta_{u,n}[t]$ is computed at local hub node (e.g. a mobile device), and $z_{u,n}[t]=1$ otherwise. Multiple tasks dispatched to the local hub node can be executed in parallel. Considering the distinct criticality of these locally-processed tasks, the local computation capability allocated to a task is proportional to the criticality index of the task. Define $V_u$ as the local computation capability of WBAN user $u$, the computation resources allocated to task $\Theta_{u,n}[t]$ is denoted as
\begin{align}
f_{u,n}^{\text{loc}}[t]=\frac{I_{u,n}[t]V_u}{\sum\limits_{n\in\mathcal{N}}(1-z_{u,n}[t])I_{u,n}[t]}.
\end{align}
Hence, the latency of local computing for task $\Theta_{u,n}[t]$ is then represented as
\begin{align}
  T^{\text{loc}}_{u,n}[t]=\frac{C_{u,n}[t]}{f_{u,n}^{\text{loc}}[t]}.
\end{align}
%*********************************************************************
\subsection{Task Offloading Model}\label{sec:offloading}
%*********************************************************************
For task offloading, both the effect of line-of-sight (LoS) and non-line-of-sight (NLoS) on wireless channel are taken into account in this work. Specifically, the probabilistic LoS model is adopted to model the large-scale attenuation between the UAV and WBAN users\cite{zeng2019energy}. The probability of geometrical LoS between the UAV and each WBAN user depends on the statistical parameters related to the environment and the elevation angle. At time slot $t$, the LoS probability for user $u$ is denoted as
\begin{align}
  \mathbb{P}^{\text{LoS}}(\beta_u[t])=\frac{1}{1+a\exp (-b(\beta_u[t]-a))},
\end{align}
where $a$ and $b$ are environment-related parameters, and $\beta_u[t]$ is the elevation angle, represented as
\begin{align}
  \beta_u[t]=\frac{180}{\pi}\arctan\left(\frac{H}{\bm{p}_u[t]-\bm{p}_v[t]}\right).
\end{align}
Then, the non-line-of-sight (NLoS) channel probability is represented as $\mathbb{P}^{\text{NLoS}}(\beta_u[t])=1-\mathbb{P}^{\text{LoS}}(\beta_u[t])$. Therefore, the expected channel gain is
\begin{align}
  g_u[t]=\frac{\mathbb{P}^{\text{LoS}}(\beta_u[t])g_0}{d_u^{\varsigma}[t]}+\frac{1-\mathbb{P}^{\text{LoS}}(\beta_u[t])\kappa g_0}{d_u^{\varsigma}[t]},
\end{align}
where $d_u^{\varsigma}[t]=\sqrt{H^2+||\bm{p}_u[t]-\bm{p}_v[t]||^2 }$ is the distance between WBAN user $u$ and the UAV at time slot $t$, $\kappa$ is the NLOS attenuation, $g_0$ is the channel gain at the reference distance $d_0$ and $\varsigma$ is the path loss exponent.

% The available system bandwidth is $W$ Hz.
To prevent the signal interference among WBAN users, the frequency bands are orthogonally allocated to users. The wireless bandwidth available for user $u$ is
$W_u$ Hz.
When delivering the tasks to the UAV for edge execution at time slot $t$, WBAN user $u$ further assigns its bandwidth and transmission power $P_{u}[t]$ to its tasks according to the task criticality index.
Let $N_0$ be the noise power at the UAV, then the transmission rate for offloading task $\Theta_{u,n}[t]$ can be obtained as
\begin{align}
    R_{u,n}[t]
   =\frac{I_{u,n}[t]W_u}{\sum\limits_{n\in\mathcal{N}}z_{u,n}[t]I_{u,n}[t]}\log_2\left(1+\frac{I_{u,n}[t]P_{u}[t]g_u[t]}{\sum\limits_{n\in\mathcal{N}}z_{u,n}[t]I_{u,n}[t]N_0}\right)
\end{align}
The data transmission time for $\Theta_{u,n}[t]$ is expressed as
\begin{align}
  T^{\text{trans}}_{u,n}[t]=\frac{D_{u,n}[t]}{R_{u,n}[t]}.
\end{align}

%*********************************************************************
\subsection{UAV Energy Consumption Model}\label{sec:UAV}
%*********************************************************************
During flight, the energy consumption of the UAV is mainly comprised of the propulsion energy consumption and the computation energy consumption. According to the existing analytical model for helicopter dynamics\cite{hu2019uav}, its propulsion energy consumption at time slot $t$ can be denoted as
\begin{align}
  E^{\text{fly}}[t]= \left(\gamma_1v^3[t]+\frac{\gamma_2}{v[t]}\right)t^{\text{fly}}.
\end{align}
where $\gamma_1$ and $\gamma_2$ are parameters related to the weight, wing area, wing span efficiency of the UAV and air density, etc.

To improve the computation energy efficiency for offloaded tasks, a dynamic voltage and frequency scaling (DVFS) technique is leveraged by the UAV. By adjusting the CPU frequency of the UAV during each time slot, its computation power can be adaptively controlled. Let $F_v$ denote as the total CPU frequency of the UAV. We consider the offloaded tasks of all WBAN users are executed concurrently by the UAV, and the computation capability allocated to an offloaded task is determined by the  proportion of its criticality index to the total criticality index of all the offloaded tasks. Thus, the CPU frequency allocated to task $\Theta_{u,n}[t]$ is
\begin{align}
  f_{u,n}^{\text{uav}}[t]=\frac{I_{u,n}[t]F_v}{\sum\limits_{u\in\mathcal{U}}\sum\limits_{n\in\mathcal{N}}{z_{u,n}[t]I_{u,n}[t]}}.
\end{align}
Then, the computation time  of task $\Theta_{u,n}[t]$ can be obtained as
\begin{align}
  T^{\text{comp}}_{u,n}[t]&=\frac{C_{u,n}[t]}{f_{u,n}^{\text{uav}}[t]}
  =\frac{C_{u,n}[t]\sum\limits_{u\in\mathcal{U}}\sum\limits_{n\in\mathcal{N}}{z_{u,n}[t]I_{u,n}[t]}}{I_{u,n}[t]F_v},
\end{align}
According to \cite{xu2017online}, the power consumption for computing task $\Theta_{u,n}[t]$ is $\eta f_{u,n}^3[t]$, where $\eta$ is the effective capacitance coefficient of the UAV, which depends on its processor chip architecture. Thus, the energy consumption for computing task $\Theta_{u,n}[t]$ is represented as
\begin{align}
  E^{\text{comp}}_{u,n}[t]=\eta (f_{u,n}^{\text{uav}}[t])^2z_{u,n}[t] C_{u,n}[t],
  =\frac{\eta z_{u,n}[t] C_{u,n}[t]I_{u,n}^2[t]F_v^2 }{\left(\sum\limits_{u\in\mathcal{U}}\sum\limits_{n\in\mathcal{N}}{z_{u,n}[t]I_{u,n}[t]}\right)^2}.
\end{align}

%*********************************************************************
\subsection{Problem Formulation}\label{sec:formulation}
%*********************************************************************
To comprehensively measure the completion time gain of tasks with different criticality in each time slot, we define a weighted task completion time for each task based on its criticality index, as follows:
\begin{align}
  \Psi_{u,n}[t]=\frac{I_{u,n}[t]}{\sum\limits_{u\in\mathcal{U}}\sum\limits_{n\in\mathcal{N}}I_{u,n}[t]}T^{\text{total}}_{u,n}[t],
\end{align}
where $T^{\text{total}}_{u,n}[t]$ is the overall completion latency of task $\Theta_{u,n}[t]$, obtained as
\begin{align}
  T^{\text{total}}_{u,n}[t]=(1-z_{u,n}[t])T^{\text{loc}}_{u,n}[t]+z_{u,n}[t](T^{\text{trans}}_{u,n}[t]+T^{\text{comp}}_{u,n}[t]).
\end{align}
 Note that it is mandatory that
each task should be completed within a time slot duration. That is, constraint $T^{\text{total}}_{u,n}[t]\le \tau$ holds.
% The established logarithmic utility expression guarantees
% that the utility of on-time task completion increases with the decrease of its completion time. With the same task completion time, the larger of the task criticality index, the higher of the utility. Besides, the logarithmic utility model  can effectively capture the  long tail effects which are close to how user experience manifests. Compared to the task completion time changes from 0.1 s to 1 s, it has less impact on
% the user experience when the task
% latency changes from 10.1 s to 11 s. Logarithmic forms
% can describe this property very well.

In this paper, we consider to minimize the weighted average task completion time of all the WBAN users during the UAV's mission period, subject to the total energy consumption of the UAV. By jointly optimizing the UAV flying trajectory and the task offloading decisions, the problem is formulated as
\begin{subequations}\label{Original}
\begin{align}\label{obj}
   \max_{v[t], \sigma[t], z_{u,n}[t]}  & \frac{1}{T}\sum\limits_{t=1}^{T}\sum\limits_{u=1}^{U}\sum\limits_{n=1}^{N}\Psi_{u,n}[t]\\
  \mbox{s.t.} \  & \sum\limits_{t=1}^{T}\left(E^{\text{fly}}[t]+\sum\limits_{u=1}^{U}\sum\limits_{n=1}^{N}E^{\text{comp}}_{u,n}[t]\right) \le E^{\text{uav}}, \label{UavEnergy}\\
  & 0 \le T^{\text{total}}_{u,n}[t] \le \tau, \forall  u\in\mathcal{U}, n\in\mathcal{N}, t\in\mathcal{T},\label{Time}\\
  & ||\bm{p}_v[t+1]-\bm{p}_v[t]||  \le V^{\max}\tau, \forall t\in\mathcal{T},\label{speed}\\
  % & \bm{p}_v[1]=(x_I,y_I)\in \mathcal{R},\bm{p}_v[t]=(x[t],y[t])\in \mathcal{R}, \label{location}\\
  & \sigma[t]\in [0,2\pi], \forall t\in\mathcal{T}, \label{angle}\\
  & z_{u,n}[t] \in \{0,1\}, \forall  u\in\mathcal{U}, n\in\mathcal{N}, t\in\mathcal{T}.\label{zt}
\end{align}
\end{subequations}
where constraint (\ref{UavEnergy}) indicates that the total energy consumption of UAV for flying and task computation is limited to its battery energy. Constraint (\ref{Time}) guarantees that task $\Theta_{u,n}[t]$ is completed within a time slot duration. The flying speed of the UAV $v[t]\in[0,V^{\max}]$ is guaranteed by constraint (\ref{speed}), where $V^{\max}$ is the maximum flight speed.
Constraint (\ref{angle}) imposes limits on the UAV's angle
of movement, and constraint (\ref{zt}) indicates the task offloading decision variables.
Problem (\ref{Original}) is a multi-stage dynamic optimization problem. Its non-convex property and complex time-correlated constraint  present
significant challenges for problem solving. Traditional optimization
algorithms often fall into the curse of dimensionality, and are hard to adapt to rapid changes
in network states. To address these issues, we propose an embodied AI framework that integrates mobility prediction and DRL to solve it in the next
section.

%*********************************************************************
\section{Hierarchical Transformer Trajectory Prediction Model}\label{sec:mobility}
%*********************************************************************
%
\begin{figure*}[t]
\centering
\includegraphics[width=1\linewidth]{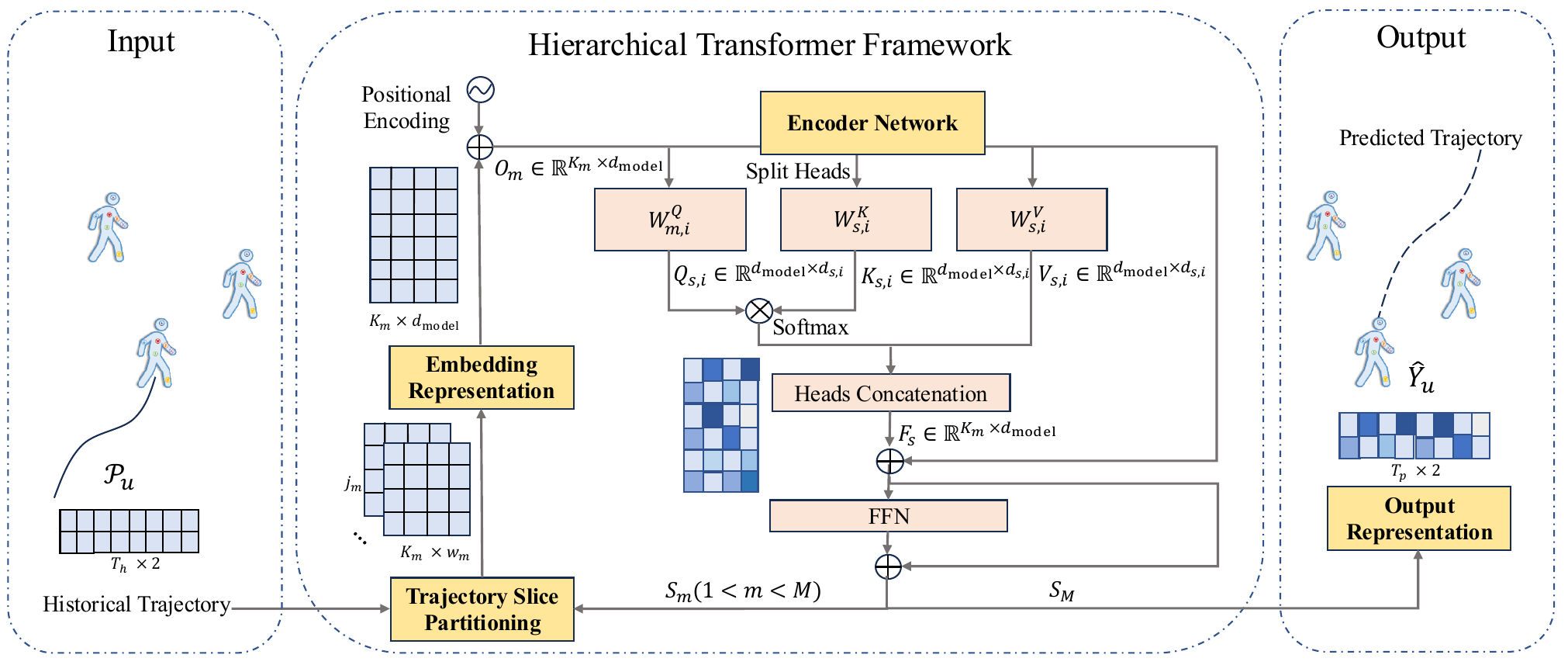}
\caption{Hierarchical Transformer Trajectory Prediction Model}
\label{transformer}
\end{figure*}
%
% Building upon the network model introduced in Section \ref{sec:model},
In this section, we propose a user trajectory prediction model based on a hierarchical multi-scale Transformer framework, to capture the temporal dependencies of user mobility on various time scales. Traditional Transformer model  \cite{vaswani2017attention} has been shown to effectively capture the long-range dependencies between words within a sentence in context of natural language processing.
 Its great sequence modeling ability facilitates to capture the contextual information of user mobility, which will help improve trajectory prediction performance. However, user trajectory typically exhibits multiple patterns with different human activities, characterized by significant
variations and fluctuations  across different
temporal scales. Traditional transformer prediction model often analyze these patterns at a unified time scale, which can lead to inaccurate learning of mobility patterns.
 Hence, we develop a hierarchical transformer framework for learning multi-scale time series features of mobility patterns.
In the following, a detailed overview of the hierarchical Transformer trajectory prediction model is provided.
% Additionally, the unifiedscale analysis approach struggles to address the non-stationary
% of mobile traffic sequences. Non-stationary in mobile traffic
% data refers to the gradual evolution of its temporal distribution
% characteristics (e.g., mean and variance) over time, posing a
% significant challenge for prediction
% Since different scale features have different receptive fields in neural networks, researchers began to extract features at different scales to improve prediction accuracy in many recent works.
% which traditional approaches struggle to model effectively.

The hierarchical Transformer trajectory prediction model is composed of four main modules: trajectory slice partitioning module, embedding representation module, encoder network module and output module. The overall structure of prediction model is illustrated in Fig. \ref{transformer}.

1) \textbf{Trajectory Slice Partitioning}: Suppose that the historical trajectory of user $u$ is represented as $\mathcal{P}_u=\{\bm{p}_u[1],\ldots,\bm{p}_u[T_h]\}$, with $T_h$ as the length of the historical observation window. The whole hierarchical Transformer framework is divided into $M$ stages that produce different feature maps of the historical trajectory for each user. To this end, a temporal slice partitioning strategy is designed to vary the time scale of the user trajectory at different stages. Specifically, a window slicing operation is leveraged to aggregate successive neighborhood location data, with the window slice size denoting the time scale size of the input.
Let the user mobility trajectory sequence at stage $m\in\mathbb{R}$ denote by $\bm{S}_{m}=[s_{m,1},s_{m,2},\ldots,s_{m,n}]$, containing $n\in\mathbb{R}$ elements with  dimensions of $j_m\in\mathbb{R}$. In particular, $\bm{S}_{1}=\mathcal{P}_u$ with $j_1=2$ denotes the raw user trajectory sequence of user $u$. The window slicing size at stage $m$ is $w_m$, which means every $w_m$ location data are grouped into a new temporal trajectory slice.
In this way, the trajectory sequence $\bm{S}_{m}$ input to stage $m$ is partitioned into a set of fine-grained trajectory slices. The number of trajectory slices $K_m$ and the size of each slice $G_m$ are respectively denoted as:
\begin{align}
  & K_m=\frac{|\bm{S}_{m}|}{w_m}, \\
  & G_m=w_m\times j_m.
\end{align}
 To some extent, it means that the length of the transformed trajectory sequence is $K_m$, and each trajectory data has feature dimension of $G_m$.

2) \textbf{Embedding Representation}:  Through the embedding layer, these trajectory slices $\bm{S}'_{m}\in\mathbb{R}^{K_m\times G_m}$ are projected into a higher dimensional space $d_{\text{model}}$. The embedding hisroty feature $\bm{Z}_m\in \mathbb{R}^{d_{\text{model}}}$ is denoted as
\begin{align}
  \bm{Z}_m = \mbox{ReLU}(\bm{S}'_{m}\bm{W^e}_m+\bm{b^e}_m),
\end{align}
where $\mbox{ReLU}(.)$\cite{goodfellow2016deep} denotes the activation function, $\bm{W^e}_m\in \mathbb{R}^{G_m\times d_{\text{model}}}$ is the embedding weight matrix and $\bm{b^e}_m\in \mathbb{R}^{K_m \times d_{\text{model}}}$ is the bias term.
To make the model understand the trajectory sequence order, locational encoding is also adopted to encode the relative locations of each data point within $\mathcal{Z}_m$. Following  \cite{vaswani2017attention}, sine and cosine functions of different frequencies are used to implement the locational encoding, shown as below:
\begin{align}
  \left\{
  \begin{array}{l}
  \text{PE}_{\text{pos},2i}=\sin\left(\frac{\text{pos}}{10000^{2i/d_{\text{model}}}}\right),  \\
  \text{PE}_{\text{pos},2i+1}=\cos\left(\frac{\text{pos}}{10000^{2i/d_{\text{model}}}}\right),
  \end{array}
\right.
\end{align}
where $\text{pos}\in\{1,2,\ldots,K_m\}$ is the location and $i\in\{1,2,\ldots,d_{\text{model}}/2\}$ is the dimension.

3) \textbf{Encoder Network}: The output $\bm{O}_m\in\mathbb{R}^{K_m\times d_{\text{model}}}$ of the locational encoding are then processed using an encoder network. Specifically, $\bm{O}_m$ is firstly transformed to a query, a key and a value through different linear projections, as the inputs of the multi-head self-attention sublayer. The query can be regarded as the transformed matrix comprised of the feature vectors of each trajectory point in $\bm{O}_m$, which is compared to the  feature vectors of every other trajectory point in the key matrix. The relevance between a query and a key is computed by dot product.
 Define the query, the key and the value of head $i$ as $\bm{Q}_{m,i}$, $\bm{K}_{m,i}$ and $\bm{V}_{m,i}$, respectively. We have
 \begin{align}
   \bm{Q}_{m,i}=\bm{O}_m\bm{W}_{m,i}^Q, \bm{K}_{m,i}=\bm{O}_m\bm{W}_{m,i}^K, \bm{V}_{m,i}=\bm{O}_m\bm{W}_{m,i}^V,
 \end{align}
 where $\bm{W}_{m,i}^Q\in \mathbb{R}^{d_{\text{model}}\times d_{i}}$, $\bm{W}_{m,i}^K\in \mathbb{R}^{d_{\text{model}}\times d_{i}}$, and $\bm{W}_{m,i}^V\in \mathbb{R}^{d_{\text{model}}\times d_{i}}$ are learnable projection parameters for head $i$ at stage $m$, respectively. $d_{i}=d_{\text{model}}/h$ is the dimension of the feature vector of head $i$ with $h$ denoting the number of heads at stage $m$.

 Then the output of the single head $i$ at stage $m$ is defined as:
 \begin{align}\label{single_head}
   \bm{A}_{m,i}=\mbox{softmax}\left(\frac{\bm{Q}_{m,i}\bm{K}_{m,i}^T}{\sqrt{d_{i}}}\right)\bm{V}_{m,i}.
 \end{align}

 Distinct attention heads are computed in parallel. Their outputs are then concatenated and projected to the dimension of $d_{\text{model}}$. The final output $\bm{F}_m\in \mathbb{R}^{K_m\times d_{\text{model}}}$ of the multi-head attention at stage $m$ is as follows:
 \begin{align}
   \bm{F}_m=[\bm{A}_{m,1},\ldots,\bm{A}_{m,h}]\bm{W^O}_m,
 \end{align}
 where $\bm{W^O}_m\in\mathbb{R}^{d_{\text{model}}\times d_{\text{model}}}$ is the weight parameter matrix of the linear projection at stage $m$.

 The output of the self-attention sublayer is subsequently fed into a feed-forward neural network (FFN) sublayer, which consists of two linear transformations with a ReLU activation in between. Define $\bm{W^{1}}_m$ , $\bm{W^{2}}_m$, $\bm{b^{1}}_m$ and $\bm{b^{2}}_m$ as learnable weights and bias of FFN at stage $m$, the process is described as:
 \begin{align}
   \bm{F'}_m=\mbox{ReLU}(\bm{F}_m\bm{W^{1}}_m+\bm{b^{1}}_m)\bm{W^{2}}_m+\bm{b^{2}}_m.
 \end{align}
 Similar to the vanilla transformer encoder, these two sublayers are then enclosed within a residual connection to form an encoder layer that avoids the vanishing gradient problem. The overall encoder network is comprised of successive encoder layers, further enhancing the model's ability to learn complex patterns and dependencies in the user trajectory. The output $\bm{F'}_m$ of the $m$-th stage is then as the input of the trajectory sequence $\bm{S}_{m+1}$ for $(m+1)$-th stage.

4) \textbf{Output Representation}: After the process of $M$ stages, the output hidden representation of the final stage $\bm{F'}_M\in \mathbb{R}^{K_M\times d_{\text{model}}}$ from encoder network is flatten, and then the predicted trajectories $\bm{Y}_u\in \mathbb{R}^{T_p\times 2}$ of user $u$ is  obtained  through a linear projection, presented as
\begin{align}
  \bm{\hat{Y}}_u = [\text{Flatten}(\bm{F'}_M)\bm{W^{x}} +\bm{b^{x}}, \text{Flatten}(\bm{F'}_M)\bm{W^{y}} +\bm{b^{y}}],
\end{align}
where $\bm{W^{x}}, \bm{W^{y}}\in \mathbb{R}^{K_Md_{\text{model}}\times T_p}$ and $\bm{b^{x}}, \bm{b^{y}}\in \mathbb{R}^{K_Md_{\text{model}}\times T_p}$ denote the output weight matrix and bias vectors for 2-D coordinates of $\bm{\hat{Y}}_u$, respectively. $T_p$ is the predicted horizon.

To evaluate the accuracy of the predicted trajectories generated by the hierarchical multi-scale Transformer, the root-mean-squared error (RMSE) is employed as the evaluation metric. RMSE measures the average Euclidean distance between predicted and ground truth locations across the prediction horizon. It is defined as
\begin{align}\label{rmse}
  \mathcal{L} =  \sqrt{\frac{1}{T_p}\sum\limits_{t=1}^{T_p}\Vert \bm{\hat{Y}}_u-\bm{Y}_u \Vert}.
\end{align}
In training phase, the hierarchical Transformer trajectory prediction model is trained by minimizing the above RMSE.
%*********************************************************************
\section{Prediction-enhanced UAV Trajectory Optimization and Task Offloading Algorithm}\label{sec:DRL}
%*********************************************************************
In this section, the UAV agent makes decisions on its flight action and task offloading decisions at each time slot with the observed and  predicted trajectory information of WBAN users.  The Markov decision process
(MDP) framework is firstly used to model problem (\ref{Original}) for the UAV edge computing network, and then a prediction-enhanced DRL algorithm for UAV trajectory optimization and task offloading is proposed.

% , a PPO-based UAV trajectory optimization and task offloading strategies is proposed.
\subsection{MDP Elements Formulation}
In this work, the key components of MDP are designed as follows.

1) \textbf{State Space}: The state space captures the environment's information at each time slot $t$. For our framework, the state encompasses the currently
observable parameters and the predicted future user trajectory information. We define the state at time slot $t$ as
\begin{align}
  s[t]=\{\bm{I}[t],\bm{P}[t],\bm{p}_v[t], E^{\text{remain}}[t]\}.
\end{align}
Here, $\bm{I}[t]=\{I_{u,n}[t]\}_{U\times N}$ is the set of task criticality index, $\bm{P}[t]=\{p_{u}[t]\}_{U\times 1}$ is the set of current user locations, $\bm{p}_v[t]$ is the UAV location and $E^{\text{remain}}[t]=E^{\text{remain}}[t-1]-\left(E^{\text{fly}}[t]+\sum_{u=1}^{U}\sum_{n=1}^{N}E^{\text{comp}}_{u,n}[t]\right)$ is  the remaining energy of the UAV at time slot $t$. Particularly, at $t = 1$, $E^{\text{remain}}[t]= E^{\text{uav}}$. Thus, the dimension of the state space is $U(N+1)+3$.

2) \textbf{Action Space}: The action space represents the decisions
made by the embodied UAV agent given a state. The selection of actions
is based on the agent隆炉s policy, which is gradually optimized throughout the learning process. Specifically, the actions include  the flying speed of the UAV, the flying angle
of the UAV, and the task offloading decisions. The action taken by the agent at time slot $t$ can be represented as
\begin{align}
a[t]=\{ v[t], \sigma[t], z_{u,n}[t]\}.
\end{align}
 Note that $v[t]\in [0,v^{\max}]$ and $\sigma[t]\in [0,2\pi]$ should be satisfied. For the task offloading decisions, we round it to the nearest integer in the range
$[0, 1]$. That is, if $z_{u,n}[t]\in [0,0.5)$, we have $z_{u,n}[t]=0$; if $z_{u,n}[t]\in [0.5, 1]$, then $z_{u,n}[t]=1$. The dimension of the action space is $UN+2$.

 3) \textbf{Reward}: The reward $r[t]$ evaluates the utility of the agent's action $a[t]$ at the given the state $s[t]$. In our algorithm, the reward function is designed to guide the UAV agent toward optimal actions by maximizing the task completion remaining time while considering the constraints on UAV energy consumption and task completion time. It is defined as
 \begin{align}
r(t)=\sum\limits_{n\in\mathcal{N}}I_{u,n}[t](\tau-T^{\text{total}}_{u,n}[t])\times \Omega^{\text{uav}} \times \Omega^{\text{time}},
 \end{align}
where $\Omega^{\text{uav}}$ and $\Omega^{\text{time}}$ are binary penalty
terms that ensure the fulfillment of constraints (\ref{UavEnergy}) and (\ref{Time}), respectively. These penalty variables are defined as follows:
\begin{align}
  \Omega^{\text{uav}}=\left\{ \begin{array}{l} 1,  \mbox{if} \ E^{\text{remain}}[t] \ge 0, \\[6pt]
   0,  \mbox{if} \ E^{\text{remain}}[t] < 0.
       \end{array} \right., \ \ \
% \end{align}
% \begin{align}
  \Omega^{\text{time}}=\left\{ \begin{array}{l} 1,  \mbox{if} \ T^{\text{total}}_{u,n}[t] \le \tau, \\[6pt]
   0,  \mbox{if} \ T^{\text{total}}_{u,n}[t] > \tau.
       \end{array} \right.
\end{align}
The reward function is strictly positive only when all
constraints in problem (\ref{Original}) are satisfied.
\subsection{Algorithm Design}
In this work, proximal policy optimization (PPO), a representative reinforcement learning
algorithm that has shown stable performance when implemented in various environments\cite{schulman2017proximal}, is utilized to implement the prediction-enhanced  UAV trajectory optimization and task offloading strategy. PPO is a policy gradient method within the actor-critic framework, where the actor network (parameterized by $\delta_A$) defines the flight and offloading policy, and the critic network (parameterized by $\delta_C$)
estimates the value function. It is improved from the trust region policy optimization (TRPO) algorithm. TRPO constrains the distance between  policies by Kullback-Leibler (KL) divergence, preventing an excessively large policy update in a single update. However, the computations using Taylor expansion approximations or conjugate gradients is overly complex. PPO, on the other hand, directly constrains the distance between the old policy $\pi_{\delta_A^{\text{old}}}$ and new policy $\pi_{\delta_A}$ within the objective function  through the clipping method. The clipped
version of the objective function is as follows:
\begin{align}\label{actor}
  \mathrm{L}^{\text{CLIP}} (\delta_A)
  =\mathbb{E}_t\left[ \min\left( \varphi_t(\delta_A) \hat{A}[t], \text{clip}(\varphi_t(\delta_A), 1 - \varepsilon, 1 + \varepsilon) \hat{A}[t] \right) \right],
\end{align}
where $\varepsilon$ is a hyper-parameter that limits the update magnitude. $\varphi_t(\delta_A)=\pi_{\delta_A}(a[t]|s[t])/\pi_{\delta_A^{\text{old}}}(a[t]|s[t])$ represents the probability ratio
between the new and old policies, and $\hat{A}[t]$ is the generalized advantage estimation at time $t$, which is calculated as
\begin{align}
  \hat{A}[t]=\sum_{l=0}^{\infty}(\gamma\lambda)^l(r[t+l]+\gamma V(s[t+l+1])-V(s[t+l])),
\end{align}
where $V(s[t])=\sum_{l=0}^{\infty}(\gamma)^lr[t+l]$ is the cumulative discounted reward, which
also represents the state-value function.
Following that, the loss function is defined as
\begin{align}\label{critic}
  &\mathrm{L}^{\text{C}}(\delta_C)=\frac{1}{2}\mathbb{E}_t\left[ (V_{\delta_C}(s[t])-V_{\text{tar}}(s[t]))^2\right],
\end{align}
where $V_{\delta_C}(s[t])$ is the value calculated by the value network
with hyper-parameters set $\delta_C$, and $V_{\delta_C}(s_{t})$ is the target value, i.e.,
\begin{align}
  V_{\text{tar}}(s[t])=r[t]+\gamma V_{\delta_C}(s[t+1]).
\end{align}
Consequently, the actor and critic can be updated according to (\ref{actor}) and (\ref{critic}),
respectively.

\begin{figure}
\centering
\includegraphics[width=5in]{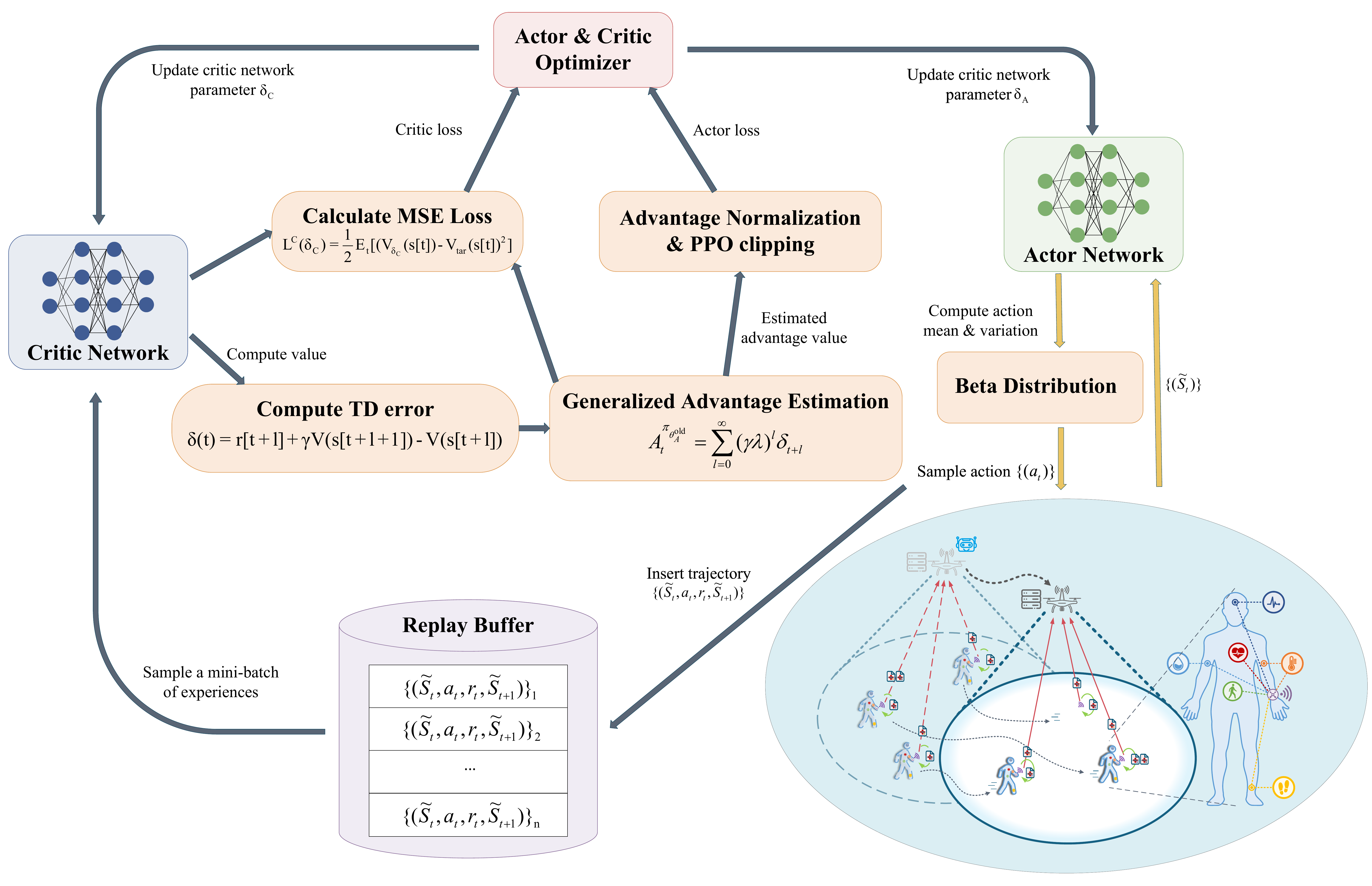}
\caption{The workflow of the interaction between the embodied UAV agent and the environment}
\label{fig:ppo}
\end{figure}

It is noted that the actions are typically
continuous and bounded in this work. Conventional action sampling from
Gaussian distribution by actor network will unavoidably
introduce an estimation bias of policy gradient, since the
boundary effects will be imposed by clipping the values
of out-of-bound actions. To tackle this problem, we adopt Beta
distribution instead of Gaussian distribution for the parameter learning of the actor network, which has the following form
\begin{align}\label{beta}
  f(x; \alpha_0, \beta_0)
  = \frac{\Gamma(\alpha_0 + \beta_0)}{\Gamma(\alpha_0)\Gamma(\beta_0)} x^{\alpha_0-1}(1-x)^{\beta_0-1}, x \in [0,1],
\end{align}
where $\alpha_0$ and $\beta_0$ are the parameters of Beta distribution. Since
(\ref{beta}) has a bounded domain, it is appropriate to sample
bounded actions.
For a clear representation of the interaction between the embodied UAV agent and environment in our proposed method, we provide the detailed workflow in Fig. \ref{fig:ppo}.

We then propose the UAV trajectory optimization and task offloading algorithm that integrates with user mobility information, shown as in Algorithm 1. In the training process of the algorithm, the UAV agent applies the trained hierarchical multi-scale Transformer model in Section \ref{sec:mobility} to predict the future mobility of WBAN users. Specifically, the UAV agent observes the current state $s[t]$ from environment at time slot $t$. Based on the historical location information $\mathcal{P}_u[t]=\{\bm{p}_u[t-T_h+1],\ldots,\bm{p}_u[t]\}$, the user mobility prediction algorithm is invoked to predict user location information $\{\bm{p}_u[t+1],\ldots,\bm{p}_u[t+T_p]\}$ for each user $u$. This information is concatenated into the environment states. The integrated states $\tilde{s}[t]$ are used for the algorithm training. Then, actions are
selected based on the actor network. The next state $s[t+1]$ is also integrated with the predicted location information to generate $\tilde{s}[t+1]$. Then,
the experience samples  are stored in the  replay buffer with the rewards feedback from environment, and the actor and critic networks
are updated using gradient descent to continuously optimize the UAV trajectory and task offloading policy.

 \begin{algorithm}[t]
 \caption{Prediction-enhanced UAV Trajectory Optimization and Task Offloading Algorithm}
 \label{drl}
 \begin{algorithmic}[1]
 \STATE \textbf{Initialization}: set the maximum episode $E^{\max}$, the maximum number of steps per episode $T^{\max}$,  discount factor $\gamma$,
hyper-parameter $\varepsilon$, learning rate, actor network $\delta_A$, critic network $\delta_C$, and create environment;
\FOR{$e=1:E^{\max}$}
  \STATE Reset the environment and obtain the initial state;
 \FOR{$t=1:T^{\max}$}
   \STATE Observe the current state $s[t]=\left\{\bm{I}[t],\bm{P}[t],\bm{p}_v[t], E^{\text{remain}}[t]\right\}$;
   \FOR{Each WBAN user $u\in \mathcal{U}$}
     \STATE Predict user location $\mathcal{P}_u[t]=\{\bm{p}_u[t+1],\ldots,\bm{p}_u[t+T_p]\}$ based on the proposed mobility prediction model;
   \ENDFOR
   \STATE Concatenate $s[t]$ with $\{\mathcal{P}_u[t]\}_U$ to obtain integrated states $\tilde{s}[t]$;
   \STATE Select action $a[t]$ using the actor network based on policy $\pi_{\delta_A}$ and get reward $r(t)$;
   \STATE Based on $s[t+1]$, predict user location $\mathcal{P}_u[t+1]=\{\bm{p}_u[t+2],\ldots,\bm{p}_u[t+T_p+1]\}$ for each user and obtain integrated states $\tilde{s}[t+1]$;
   \STATE Record the transition tuple $(\tilde{s}[t],a[t],r[t],\tilde{s}[t+1])$ into the experience replay buffer;
   \STATE Sample random mini-batch of transitions from the experience replay buffer;
   \STATE Update actor network $\delta_A$ via gradient descent on $\mathrm{L}^{\text{CLIP}} (\delta_A)$;
   \STATE Update critic network $\delta_C$ by minimizing $\mathrm{L}^{\text{C}} (\delta_C)$ ;
  \STATE Update the state of the environment.
 \ENDFOR
 \ENDFOR
 \end{algorithmic}
 \end{algorithm}
 %*********************************************************************
 \section{Performance Evaluation}\label{sec:eva}
 %*********************************************************************
 This section presents a comprehensive evaluation of the proposed hierarchical Transformer trajectory prediction model and the PETO algorithm. Specifically, we introduces the experimental dataset and simulation setting in subsection \ref{sec:setting}. Then, we
conduct testing and validation to assess the performance of the trajectory prediction model in subsection \ref{sec:PreEva}, while subsection \ref{sec:ToEva} presents the simulation results for PETO algorithm in terms of weighted task completion time compared to baseline methods.

 %*********************************************************************
 \subsection{Experiment Setting}\label{sec:setting}
 %*********************************************************************
 \begin{table}[t]
 \centering
 \caption{Environmental Parameter Settings}
 \begin{tabular}{ccc}
 \toprule
 Description &  \makecell{Value} \\ \toprule
 Altitude of UAV & 100m\cite{tang2025deep}\\
 \specialrule{0em}{2pt}{1pt}
 Mission Period $T^{\max}$ & 100s \cite{liu2023energy}\\
 \specialrule{0em}{2pt}{1pt}
 Data load $D_{u,n}[t]$ & [1,2]MB\\
 \specialrule{0em}{2pt}{1pt}
 Computation amount $C_{u,n}[t]$ & [1,2] Gigacycles\\
 \specialrule{0em}{2pt}{1pt}
 Local computation capability $V_u$ of user $u$ &  1 Gigacycles/s\cite{liu2023energy}\\
 \specialrule{0em}{2pt}{1pt}
 Edge computation capability $F_v$ of UAV $d_{\text{model}}$  & 10 Gigacycles/s\cite{liu2023energy}\\
 \specialrule{0em}{2pt}{1pt}
 Channel bandwidth $W_u$ for user $u$ &  1MHz \\
 \specialrule{0em}{2pt}{1pt}
 Parameters of LOS channel $a$,$b$ & 10, 0.6\cite{zeng2019energy}\\
 \specialrule{0em}{2pt}{1pt}
 NLOS attenuation $\kappa$ & 0.2\cite{yang2022online}\\
 \specialrule{0em}{2pt}{1pt}
 Path loss exponent $\varsigma$ & 2.3\cite{yang2022online} \\
 \specialrule{0em}{2pt}{1pt}
 Channel gain at reference distance $g_0$ & 1.42e-4\cite{wang2021deep}\\
 \specialrule{0em}{2pt}{1pt}
 Initial Energy $E^{\text{UAV}}$ of UAV & 500kJ\cite{chen2025computation}\\
 \specialrule{0em}{2pt}{1pt}
 Effective capacitance coefficient $\eta$ & 1e-27\cite{liu2023energy} \\
 \specialrule{0em}{2pt}{1pt}
 Maximum flying speed $V^{\max}$ & 50m/s\cite{chen2025computation}  \\
 \specialrule{0em}{2pt}{1pt}
 Transmission power $P_u[t]$ & 100mW\cite{ren2024resource}  \\
 \specialrule{0em}{2pt}{1pt}
 Noise Power $N_0$ & -60dBm\cite{hu2019uav}  \\
 \specialrule{0em}{2pt}{1pt}
 \bottomrule
 \end{tabular}
 \label{table.parameters1}
 \end{table}
To evaluate the performance of the proposed trajectory prediction model, a
real-world public human trajectory dataset is used for training and testing.
The trajectory dataset was collected in the GeoLife project by Microsoft Research Asia\cite{zheng2010geolife}. The dataset contains precise latitude and longitude information on the consecutive locations of 182 users obtained from the GPS timestamp. The frame rate we adopt is 1s. For the trajectory data with intermittent missing values,
we apply linear interpolation between adjacent observations. Each user trajectory is split into multiple
segments with duration of $T_h+T_p$ seconds by window sliding, and use the data of last $T_h$ seconds
to predict the user's trajectory in the next $T_p$ seconds. For training, validation and testing, the data set is divided in a 7:2:1 ratio.

\begin{table}[t]
\centering
\caption{Parameters in the Proposed Methods}
\begin{tabular}{ccc}
\toprule
Description &  \makecell{Value} \\ \toprule
Length of historical trajectory $T_h$ & 60s \\
\specialrule{0em}{2pt}{1pt}
Length of predicted trajectory $T_p$ & 10s \\
\specialrule{0em}{2pt}{1pt}
Number of stages $M$ & 3 \\
\specialrule{0em}{2pt}{1pt}
Trajectory slice size $w_m$ &  2\\
\specialrule{0em}{2pt}{1pt}
Hidden size of the output $d_{\text{model}}$  & 64 \\
\specialrule{0em}{2pt}{1pt}
Number of encoders &  6 \\
\specialrule{0em}{2pt}{1pt}
Batch size  & 64  \\
\specialrule{0em}{2pt}{1pt}
Learning rate & 1e-3  \\
\specialrule{0em}{2pt}{1pt}
Clipping parameter $\varepsilon$ & 0.2  \\
\specialrule{0em}{2pt}{1pt}
Hidden size of the actor/critic network & 128 \\
\specialrule{0em}{2pt}{1pt}
Discount factor $\gamma$ & 0.98  \\
\specialrule{0em}{2pt}{1pt}
Replay buffer size & 1e6  \\
\specialrule{0em}{2pt}{1pt}
\bottomrule
\end{tabular}
\label{table.parameters2}
\end{table}
For the performance evaluation on PETO algorithm, we then consider an environment  containing 10 mobile WBAN users that generate tasks with different criticality. According to the 802.15.6 protocol\cite{kwak2010overview}, the services provided by the WBAN can be divided into four categories: non-medical services, low-priority medical services, general health services, and high-priority medical services. Each BAN user is equipped with five different physiological sensors to monitor data with different criticality, including background information, voice, ECG, body temperature, and movement, and generates corresponding data analysis tasks. Therefore, $\phi_u\in [0.25,0.5,0.75,1]$, and  $\rho_{u,n}\in[0.2,0.4,0.6,0.8,1]$ after normalization. The urgency of the perceived data is divided into normal
data and emergency abnormal data, i.e., $\alpha_{u,n}\in[0.5,1]$. Criticality index $I_{u,n}[t]$ is defined as $(\phi_u + \rho_{u,n} +\alpha_{u,n})/3$. The criticality of the data at different time slots follows the Markov property, and the corresponding state transition probability matrix is $[0.7,0.3;0.3,0.7]$\cite{yuan2018performance}. To facilitate the distance computation between UAV and users, user locations are converted from the World Geodetic System coordinates to the Cartesian coordinate system by Haversine formula.
 Each time slot
is divided into 1s of movement time and 1s of offloading
and computation time\cite{hu2024drl}. Unless
otherwise stated, the detailed environmental
parameters are listed in TABLE \ref{table.parameters1}.
For the implementation details of our hierarchical Transformer trajectory prediction model and the PETO algorithm, the default parameters are listed in TABLE \ref{table.parameters2}. Adam optimizer and standard normalization are employed for model training in both of the methods. To mitigate overfitting, we employ early stopping during training. Specifically, we monitor the validation loss and halt the training process if it does not decrease for 10 consecutive epochs, restoring the model parameters that achieved the best validation performance. In addition, all the experiments are implemented with PyTorch 1.13 \cite{paszke2019pytorch} with Python 3.8 and trained on a T4 with 2560 CUDA cores.

%*********************************************************************
\subsection{Evaluation of Trajectory Prediction Model}\label{sec:PreEva}
%*********************************************************************
%
\begin{figure}[t]
\centering
\includegraphics[width=3.5in]{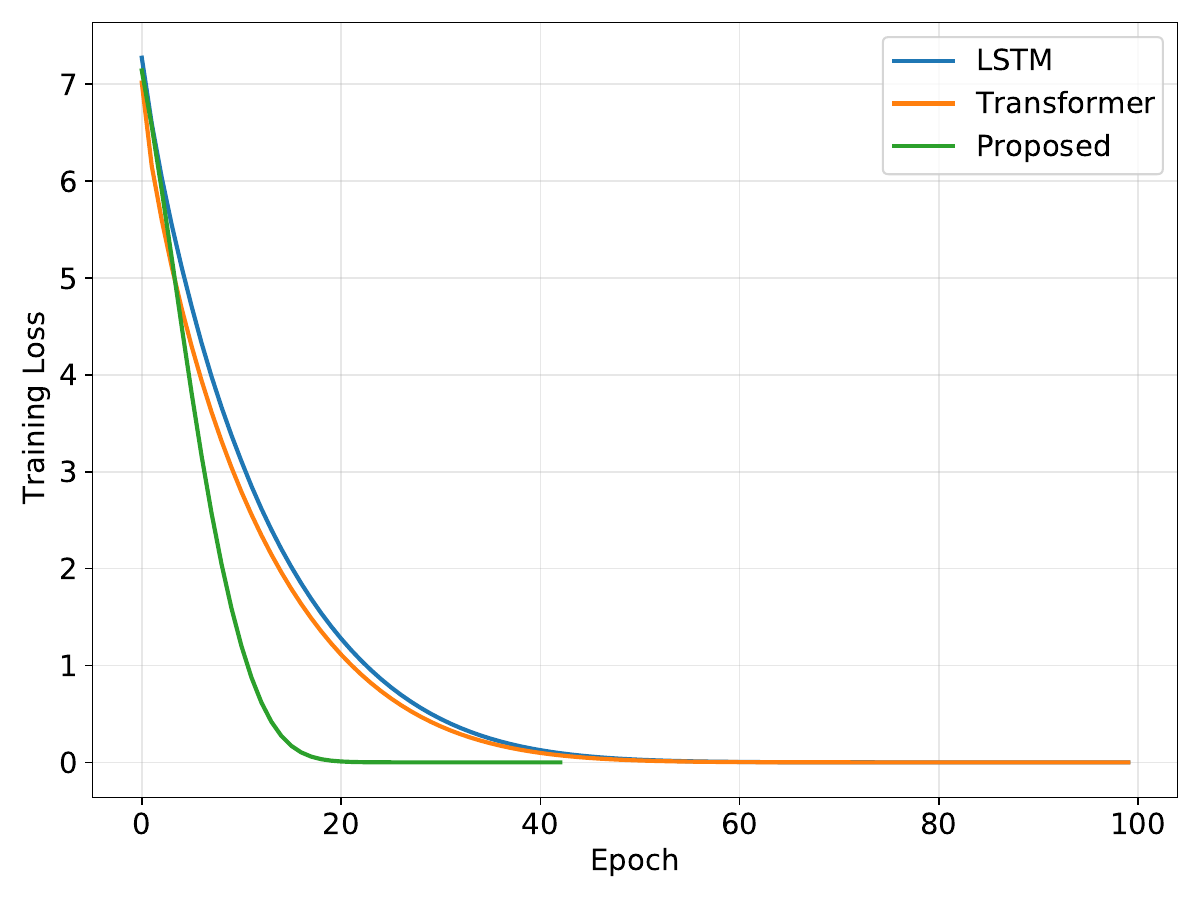}
\caption{Convergence behavior of mobility prediction models}
\label{fig:prediction_conv}
\end{figure}

To verify the effectiveness of the proposed trajectory prediction model, two prediction methods in the existing work  are also implemented for comparative analysis. Specifically, LSTM-based user trajectory prediction\cite{ma2020leveraging} serves as the baseline algorithm. In the model, the LSTM cell with size 128 is used. Following that, two fully connected layers with the activation functions ReLU are added. The vanilla Transformer is also utilized to  predict the  user trajectory \cite{najjar2024pre}, in which the number of encoders and the attention heads are the same to the proposed hierarchical Transformer trajectory prediction model.
These algorithms are repeated for 10 times with different seeds. The mean value of RMSE calculated using (\ref{rmse}) from 10 experiments are reported.

Fig. \ref{fig:prediction_conv} shows the convergence behavior of the three trajectory prediction models, in which their training loss are plotted with epochs. As epochs increases, the training loss of the three trajectory prediction models gradually decreases and approximately converges to 0. It can be observed that the proposed hierarchical Transformer trajectory prediction model exhibits more rapid convergence compared to the other models, reaching a stable state approximately 20 epochs with early stopping triggered at epoch 42 to prevent overfitting.

\begin{figure}[t]
\centering
\includegraphics[width=3.5in]{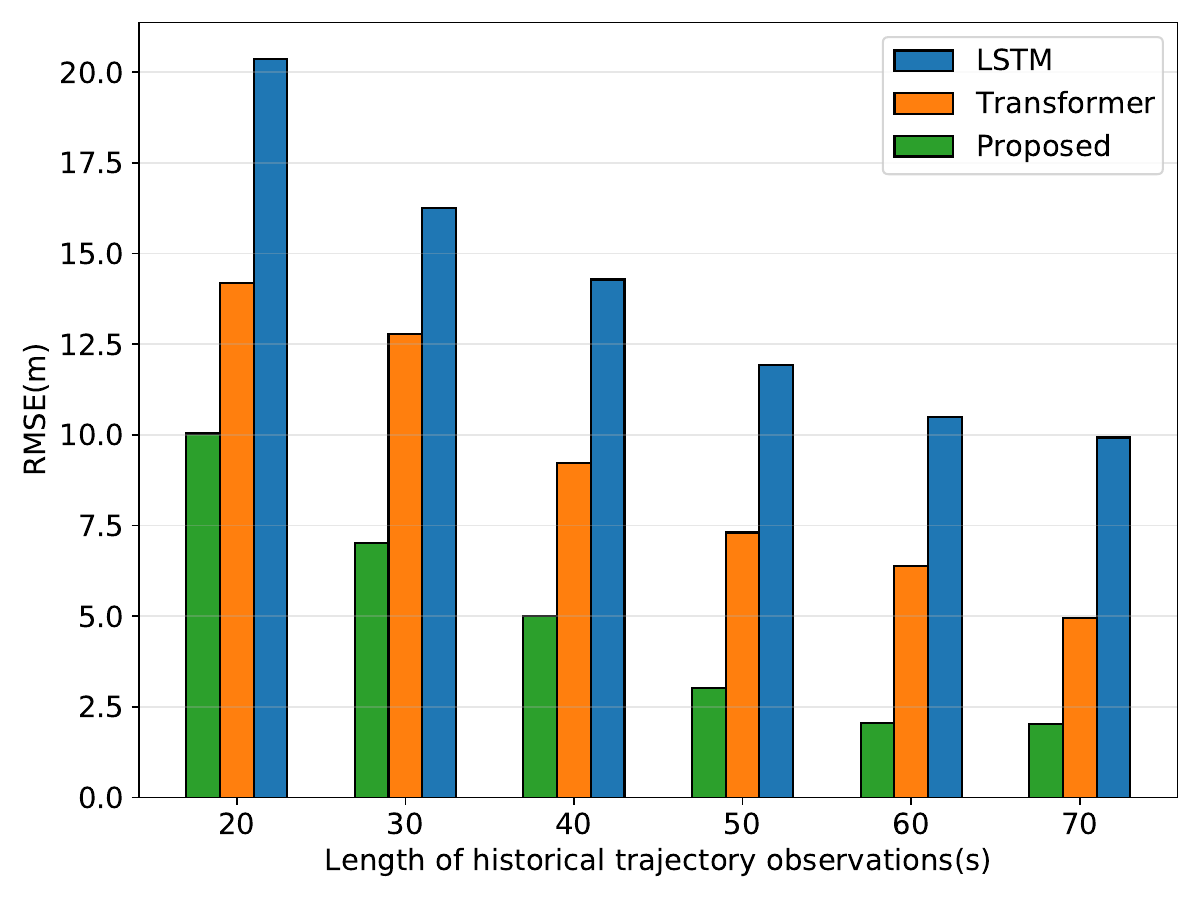}
\caption{RMSE comparison for user trajectory prediction models}
\label{fig:RMSE_comparison}
\end{figure}
%
% \begin{table*}[t]
% \centering
% \caption{RMSE comparison for user trajectory prediction models}
% \setlength{\tabcolsep}{1.5mm}
% \begin{tabular}{ccccccc}
% \toprule
% \multirow{2}{*}{Models} & \multicolumn{6}{c}{Length of historical trajectory observations}   \\ \cline{2-7}
%  \specialrule{0em}{2pt}{1pt}
%   &20s & 30s & 40s & 50s & 60s & 70s \\ \toprule
%  \specialrule{0em}{2pt}{1pt}
%  LSTM & 20.352 $\pm$ 0.015 & 16.266 $\pm$ 0.011 & 14.285 $\pm$ 0.012  & 11.936 $\pm$ 0.008 & 10.487 $\pm$ 0.017 & 9.927 $\pm$ 0.006\\
% \specialrule{0em}{2pt}{1pt}
% Transformer & 14.198 $\pm$ 0.010 & 12.782 $\pm$ 0.015 &9.226 $\pm$ 0.025 & 7.309 $\pm$ 0.047 & 6.382 $\pm$ 0.033 & 4.952 $\pm$ 0.017\\
% \specialrule{0em}{2pt}{1pt}
% Proposed & 10.037 $\pm$ 0.005 & 7.018 $\pm$ 0.002 & 5.015 $\pm$ 0.003 & 3.024 $\pm$ 0.006 & 2.053 $\pm$ 0.002 & 2.023 $\pm$ 0.007 \\
% \specialrule{0em}{2pt}{1pt}
% \bottomrule
% \end{tabular}
% \label{table.rmse_results}
% \end{table*}

To verify the prediction performance of different user trajectory prediction models, Fig. \ref{fig:RMSE_comparison} presents their RMSE results on predicted trajectories against the actual ones over the historical observations from 20s to 70s. The predicted horizon is set to 10s. As shown in the figure, a longer historical observation windows contributes to more accurate predictions, in spite of the higher complexity. The improvement gap become small with the increasing length of historical observations. For the proposed hierarchical Transformer trajectory prediction model, the historical observations of 60s are enough for a good balance between accuracy and complexity.
 Compared to the LSTM model, both the two Transformer-based models achieve lower RMSE, particularly when the historical observation is long. Notably, the proposed hierarchical Transformer trajectory prediction model reduces RMSE by an average of 67.86\%, with the most pronounced drop of 80.42\% at 60s. This suggests that the multi-head self-attention attention mechanism in vanilla Transformer model and the proposed model can better capture long-range temporal dependencies in the user trajectory, which provides better mobility prediction.
Besides, it can be observed that the  proposed trajectory prediction model outperforms the vanilla Transformer model with fixed-scale feature representation, showing a consistent RMSE reduction with an average of 46.82\%. This demonstrates the hierarchical feature extraction from small-scale fine-grained temporal trajectory features to large-scale coarse-grained trajectory features  is more effective for prediction.

\begin{figure}[t]
\centering
\includegraphics[width=3.5in]{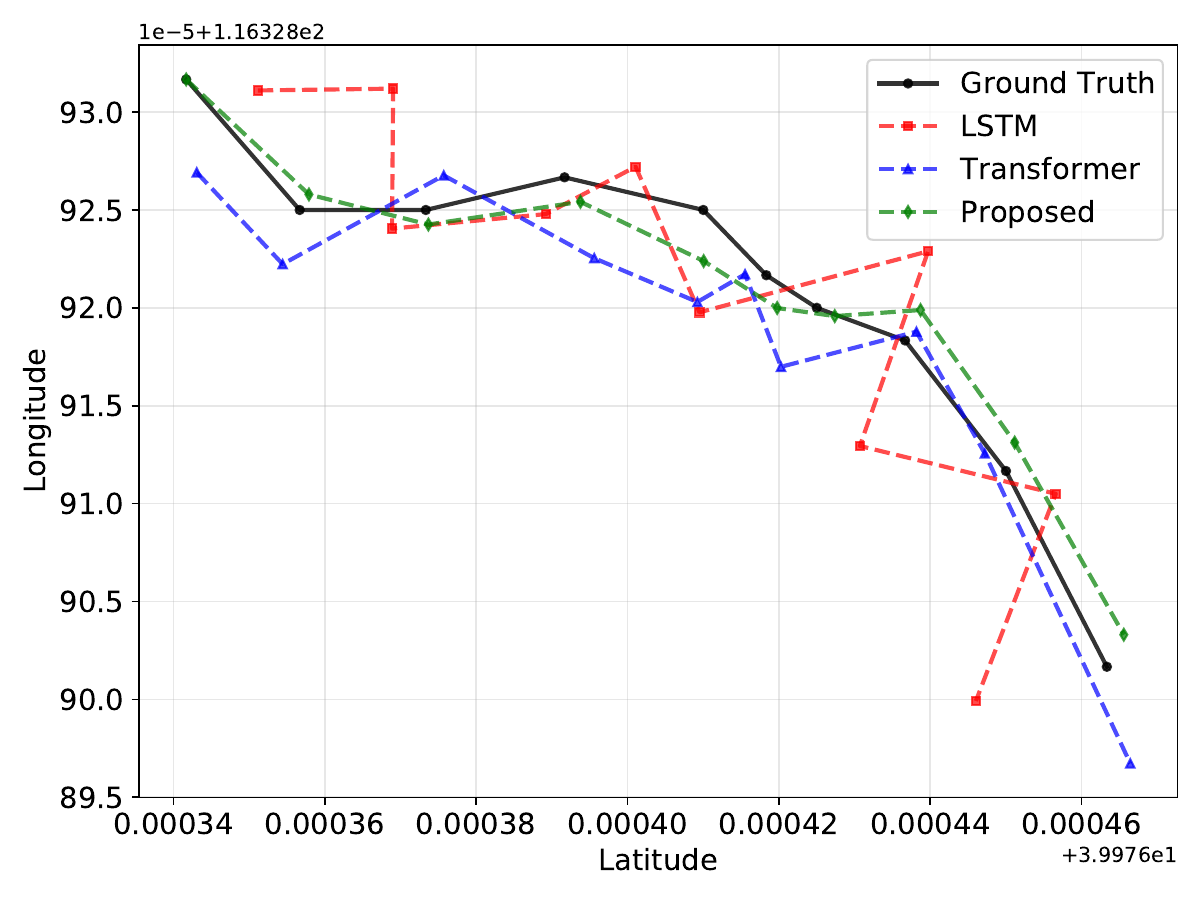}
\caption{Trajectory prediction results of different models}
\label{fig:trajectory}
\end{figure}

In addition to quantitative metrics, the trajectory prediction results of different models are visualized in Fig. \ref{fig:trajectory}. In the figure, the prediction results  of a representative user (ID 20) is illustrated. The X-axis represents the latitude value, and the y-axis represents the longitude value. Both the units of the X-axis and Y-axis are decimal degree of World Geodetic System coordinates. It is noted that the trajectory obtained by the proposed hierarchical Transformer prediction model is closely aligned with the actual trajectory, which exhibits its superior fitting accuracy. This result highlights the model's ability to capture complex temporal dependencies of user movements.
%

%*********************************************************************
\subsection{Evaluation of PETO Algorithm}\label{sec:ToEva}
%*********************************************************************
In this subsection, we validate the performance of the proposed PETO algorithm. The convergence results and the optimized UAV trajectory are firstly provided. Then, we compare the PETO algorithm with the following benchmark and state-of-the-art methods:
\begin{itemize}
  \item[1)] Random UAV trajectory and edge computing (RUEC): At each time slot, the UAV agent randomly chooses the flying speed and angle, and all the tasks are offloaded to the UAV by each WBAN user. Each task $\Theta_{u,n}[t]$ is allocated computation resources according to their criticality index $I_{u,n}[t]$.

  \item[2)] PPO-based UAV trajectory optimization and task offloading algorithm without prediction (PAWP) \cite{wang2025joint}: At each time slot, the UAV agent observes the current system state, based on which the UAV agent optimizes trajectory and makes task offloading decisions by PPO reinforcement learning algorithm.

  \item[3)] DDPG-based UAV trajectory optimization and task offloading algorithm with prediction (DAWP) \cite{wang2021computation}:  At each time slot, the UAV agent utilizes the proposed trajectory model to predict the future user trajectory. According to the current state and the predicted partial state, the UAV agent optimizes trajectory and makes task offloading decisions by DDPG reinforcement learning algorithm.
\end{itemize}
\begin{figure}[t]
\centering
\includegraphics[width=3.5in]{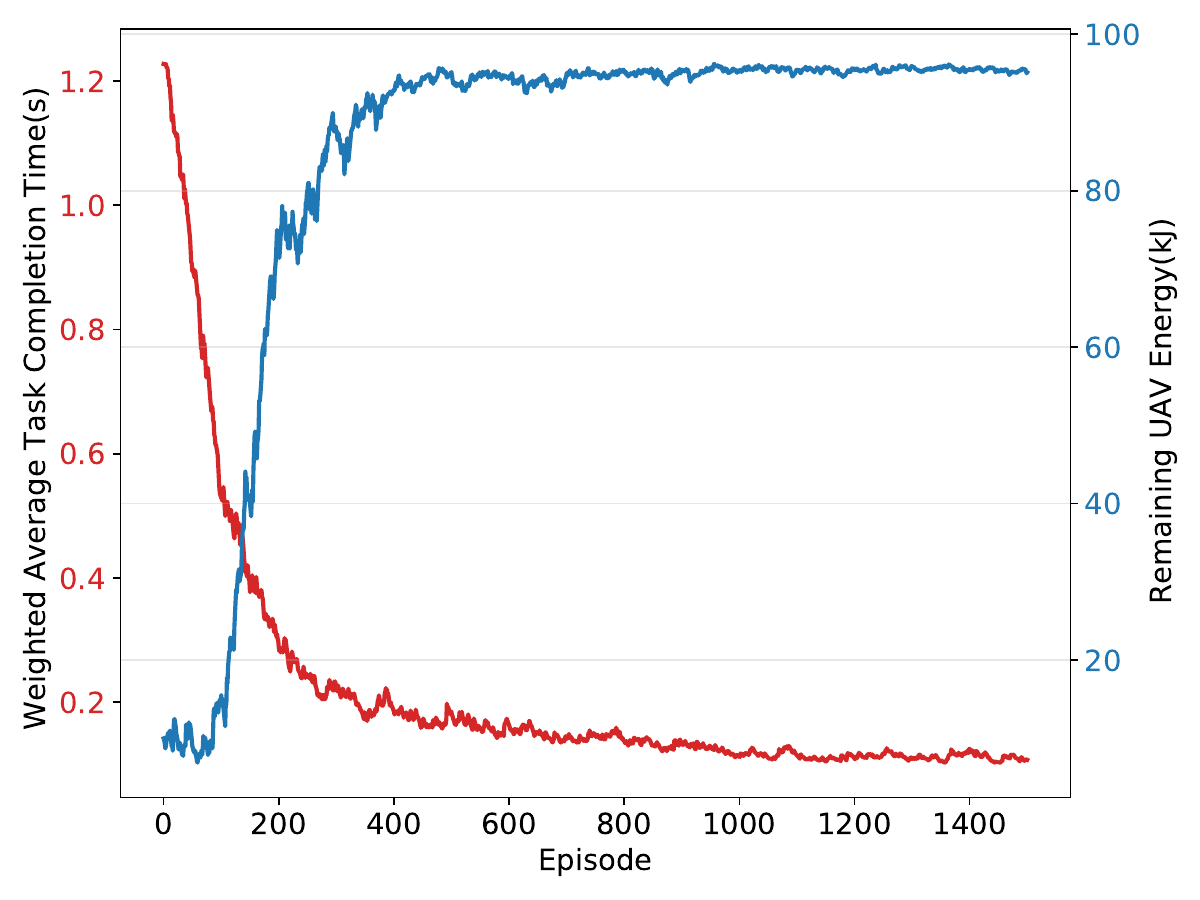}
\caption{Convergence behavior of the proposed PETO algorithm}
\label{fig:convergence}
\end{figure}
\begin{figure}[t]
\centering
\includegraphics[width=3.5in]{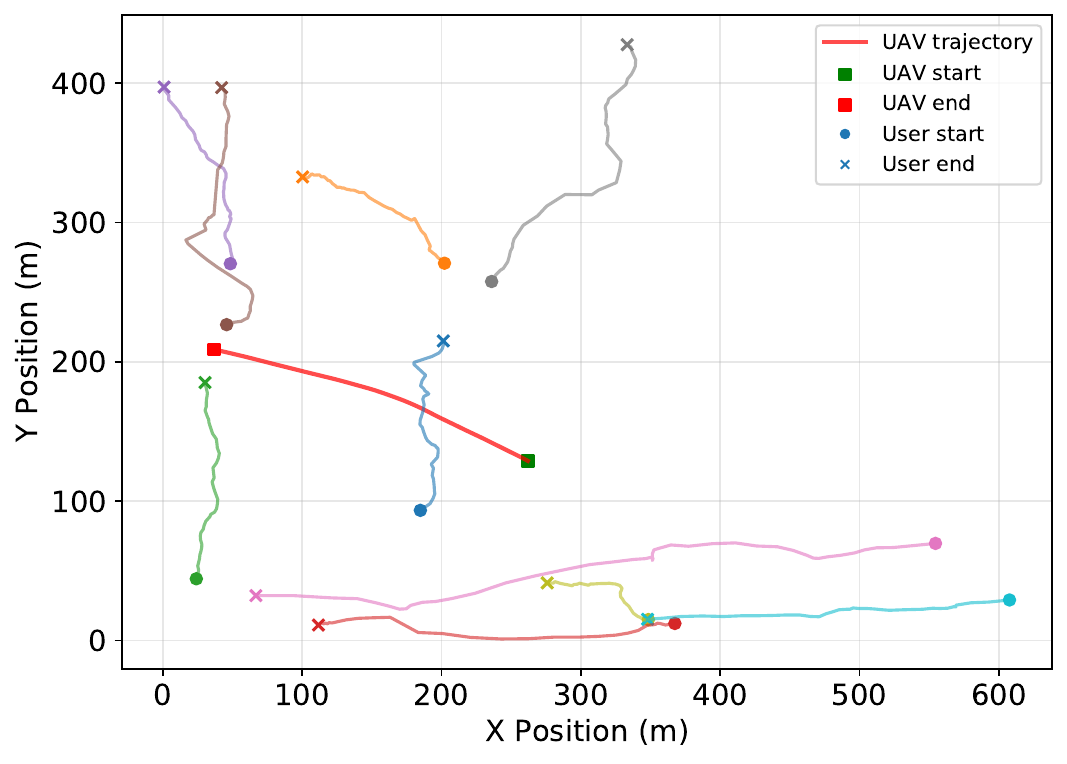}
\caption{UAV trajectory of the proposed PETO algorithm}
\label{fig:uav_trajectory}
\end{figure}
\begin{figure}[t]
\centering
\includegraphics[width=3.5in]{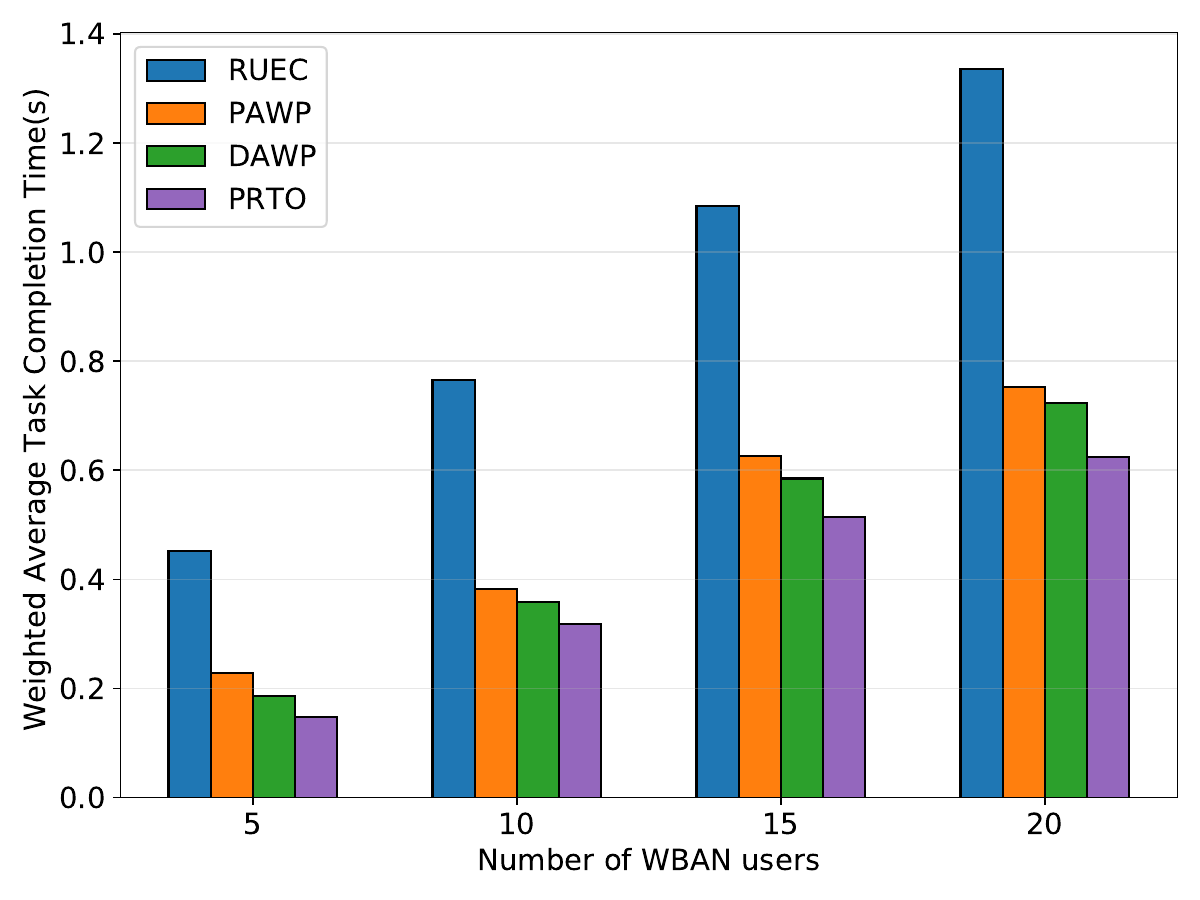}
\caption{Weighted average task completion time versus number of WBAN users}
\label{fig:alg_compr_rl}
\end{figure}

Fig. \ref{fig:convergence} shows the convergence behavior of our PRTO algorithm, in which the weighted average task completion time and the UAV remaining energy in each episode are plotted with the evolution of episodes. In the early training stage, such as 0-1000 episodes, the PRTO algorithm still maintains a high exploration rate, causing the UAV agent to frequently try sub-optimal actions. As episode evolves, the weighted average task completion time declines and the UAV remaining energy grows as better trajectory and offloading strategies are learned from the environment dynamics, leading to more satisfactory performance.  The results indicate that our PRTO
method  has stable convergence behavior and is reliable in this dynamic network environment. In addition, Fig. \ref{fig:uav_trajectory}  characterizes both the users' trajectory and the designed UAV trajectory by the proposed PETO algorithm. The initial position of the UAV is determined by the geometric center of the user locations. It can be observed that the UAV agent intelligently follows a learned optimal trajectory in response to the movements of WBAN users, thereby maintaining the quality of edge computing services.

In Fig. \ref{fig:alg_compr_rl}, the performance comparison on weighted average task completion time of different solutions are illustrated as the number of WBAN users varies. With the increasing number of WBAN users, more users compete the limited computation resources of UAV, and the computation resources allocated to each task decrease. As a result, the weighted average task completion time grows in all the four methods. Note that  the increase in weighted average task completion time gradually become small for the PAWP, DAWP and the proposed PRTO methods, this is because the computational demands of tasks from more WBAN users (e.g., 20) exceed the afforded computational capability of the UAV, results in most of the tasks being processed locally.
It is observed the proposed PRTO method consistently achieves superior performance compared to the other methods. Specifically, the proposed PRTO achieves 57.95\% decrease in weighted average task completion time over RUEC on average. The performance gain comes from better flight trajectory and offloading decisions through the interactive learning of the embodied UAV agent. Without parallel computing at the local device and the UAV, too much tasks from WBAN users overloads the UAV in the RUEC method, leading to longer average task completion time.  Compared to the PAWP method, the proposed PRTO method demonstrates approximately 21.76\% gain in weighted average task completion time reduction. This improvement is attributed to the advanced prediction of user mobility by the proposed hierarchical Transformer framework, which facilitates the UAV to relocate a better position at each time slot to provide the edge computing service.
 Compared to DAWP, PRTO method also has better performance, which benefits from the use of
generalized advantage estimation in PRTO provides more accurate advantage estimation and more effective action learning. These results demonstrate the scalability of our
PRTO method, as it maintains superior performance
gains even as the network size grows.

% Fig. \ref{} shows the convergence comparison of our PRTO algorithm and the other three baselines. in which the total rewards of all the steps in each episode are plotted with the evolution of episodes. In the early training stage, such as 0-40 episodes, the PRTO algorithm still maintains a high exploration rate, causing the UAV agent to frequently try sub-optimal actions, leading to the fluctuated rewards. The total rewards increase rapidly as better trajectory and offloading strategy are learned from the environment dynamics.  The results indicate that our PRTO
% method converges to superior performance compared to the other methods.  The convergence results demonstrates the proposed PRTO algorithm is more reliable in this dynamic network environments. have more stable
% convergence behavior
\section{Conclusions}\label{sec:con}
%**********************************************************************************
In this paper, we have proposed an embodied AI-enhanced IoMT edge computing framework, in which the UAV embodied AI agent serves as the edge server for mobile WBAN users with time-varying task criticality.
By integrating the designed hierarchical Transformer model for user mobility prediction and DRL for UAV flight trajectory and task offloading decision-making, the proposed method can reduce task completion time and improve adaptability in dynamic IoMT environments.
Real-word traces have demonstrated that our proposed user mobility prediction method consistently
outperforms traditional methods in terms of convergence
speed and the prediction accuracy. In addition, simulations results have shown the proposed DRL algorithm greatly improves the task completion time performance through prediction enhancement, validating the effectiveness of using embodied AI framework in IoMT edge computing scenario.

\bibliographystyle{IEEEtran}
\bibliography{IEEEabrv,MCO}

@article{zhu2022iomt,
  title={{IoMT}-enabled real-time blood glucose prediction with deep learning and edge computing},
  author={Zhu, Taiyu and Kuang, Lei and Daniels, John and Herrero, Pau and Li, Kezhi and Georgiou, Pantelis},
  journal={IEEE Internet Things J.},
  volume={10},
  number={5},
  pages={3706--3719},
  year={2022},
  publisher={IEEE}
}

@article{wu2023mobility,
  title={Mobility-aware deep reinforcement learning with seq2seq mobility prediction for offloading and allocation in edge computing},
  author={Wu, Chao-Lun and Chiu, Te-Chuan and Wang, Chih-Yu},
  journal={IEEE Trans. Mob. Comput.},
  volume={23},
  number={6},
  pages={6803--6819},
  year={2023},
  publisher={IEEE}
}

@article{ma2020leveraging,
  title={Leveraging the power of prediction: Predictive service placement for latency-sensitive mobile edge computing},
  author={Ma, Huirong and Zhou, Zhi and Chen, Xu},
  journal={IEEE Trans. Wireless Commun.},
  volume={19},
  number={10},
  pages={6454--6468},
  year={2020},
  publisher={IEEE}
}

@inproceedings{najjar2024pre,
  title={Pre-trained Transformer uncovers meaningful patterns in human mobility data},
  author={Najjar, Alameen},
  booktitle={IEEE Int. Conf. Big Data (BigData)},
  pages={5819--5828},
  year={2024},
}

@article{zhang2024generative,
  title={Generative AI agents with large language model for satellite networks via a mixture of experts transmission},
  author={Zhang, Ruichen and Du, Hongyang and Liu, Yinqiu and Niyato, Dusit and Kang, Jiawen and Xiong, Zehui and Jamalipour, Abbas and Kim, Dong In},
  journal={IEEE J. Sel. Areas Commun.},
volume={42},
  number={12},
  pages={3581-3596},
  year={2024},
  publisher={IEEE}
}

@article{zhang2025embodied,
  title={Embodied {AI}-enhanced vehicular networks: An integrated vision language models and reinforcement learning method},
  author={Zhang, Ruichen and Zhao, Changyuan and Du, Hongyang and Niyato, Dusit and Wang, Jiacheng and Sawadsitang, Suttinee and Shen, Xuemin and Kim, Dong In},
  journal={IEEE Trans. Mob. Comput.},
	volume={24},
  number={11},
  pages={11494--11510},
  year={2025},
  publisher={IEEE}
}

@article{mu2025aoi,
  title={AoI-Aware Online Transmission Optimization for {WBANs} with Unreliable Information Delivery},
  author={Mu, Siqi and Lu, Yang and Jiang, Ruihong and Chen, Wei and Ai, Bo and Niyato, Dusit},
  journal={IEEE Trans. Wireless Commun.},
  year={DOI:10.1109/TWC.2025.3620434, 2025},
  publisher={IEEE}
}

@inproceedings{guo2019joint,
  title={Joint trajectory and computation offloading optimization for {UAV}-assisted {MEC} with {NOMA}},
  author={Guo, Fengxian and Zhang, Heli and Ji, Hong and Li, Xi and Leung, Victor CM},
  booktitle={IEEE Conf. Comput. Commun. Workshops(INFOCOM WKSHPS)},
  pages={1--6},
  year={2019},
}

@article{liu2023maximizing,
  title={Maximizing energy efficiency in {UAV}-assisted {{NOMA}-{MEC}} networks},
  author={Liu, Zhixin and Qi, Junxiao and Shen, Yanyan and Ma, Kai and Guan, Xinping},
  journal={IEEE Internet Things J.},
  volume={10},
  number={24},
  pages={22208--22222},
  year={2023},
  publisher={IEEE}
}

@article{rahman2021internet,
  title={An {Internet}-of-medical-things-enabled edge computing framework for tackling {COVID-19}},
  author={Rahman, Md Abdur and Hossain, M Shamim},
  journal={IEEE Internet Things J.},
  volume={8},
  number={21},
  pages={15847--15854},
  year={2021},
  publisher={IEEE}
}

@article{ning2020mobile,
  title={Mobile edge computing enabled 5G health monitoring for {Internet} of medical things: A decentralized game theoretic approach},
  author={Ning, Zhaolong and Dong, Peiran and Wang, Xiaojie and Hu, Xiping and Guo, Lei and Hu, Bin and Guo, Yi and Qiu, Tie and Kwok, Ricky YK},
  journal={IEEE J. Sel. Areas Commun.},
  volume={39},
  number={2},
  pages={463--478},
  year={2020},
  publisher={IEEE}
}

@article{vaswani2017attention,
  title={Attention is all you need},
  author={Vaswani, Ashish and Shazeer, Noam and Parmar, Niki and Uszkoreit, Jakob and Jones, Llion and Gomez, Aidan N and Kaiser, {\L}ukasz and Polosukhin, Illia},
  journal={Adv. Neural Inf. Process. Syst.},
  volume={30},
  year={2017}
}

@book{goodfellow2016deep,
  title={Deep learning},
  author={Goodfellow, Ian and Bengio, Yoshua and Courville, Aaron},
  year={2016},
  publisher={MIT press}
}

@article{schulman2017proximal,
  title={Proximal policy optimization algorithms},
  author={Schulman, John and Wolski, Filip and Dhariwal, Prafulla and Radford, Alec and Klimov, Oleg},
  journal={arXiv preprint arXiv:1707.06347},
  year={2017}
}

@article{zheng2010geolife,
  title={GeoLife: A collaborative social networking service among user, location and trajectory.},
  author={Zheng, Yu and Xie, Xing and Ma, Wei-Ying and others},
  journal={IEEE Data Eng. Bull.},
  volume={33},
  number={2},
  pages={32--39},
  year={2010}
}

@inproceedings{kwak2010overview,
  title={An overview of IEEE 802.15.6 standard},
  author={Kwak, Kyung Sup and Ullah, Sana and Ullah, Niamat},
  booktitle={Int. Symp. Appl. Sci. Biomed. Commun. Technol. (ISABEL)},
  pages={1--6},
  year={2010},
}

@article{yuan2018performance,
  title={Performance analysis of IEEE 802.15.6-based coexisting mobile {WBANs} with prioritized traffic and dynamic interference},
  author={Yuan, Xiaoming and Li, Changle and Ye, Qiang and Zhang, Kuan and Cheng, Nan and Zhang, Ning and Shen, Xuemin},
  journal={IEEE Trans. Wireless Commun.},
  volume={17},
  number={8},
  pages={5637--5652},
  year={2018},
  publisher={IEEE}
}

@article{hu2024drl,
  title={{DRL}-based trajectory optimization and task offloading in hierarchical aerial {MEC}},
  author={Hu, Zhihao and Yang, Yaozong and Gu, Wei and Chen, Ying and Huang, Jiwei},
  journal={IEEE Internet Things J.},
  year={2024},
	volume={12},
	number={3},
	pages={3410--3423},
  publisher={IEEE}
}

@article{wang2023joint,
  title={Joint trajectory design and power allocation for {UAV} assisted network with user mobility},
  author={Wang, Jing and Zhou, Xiaotian and Zhang, Haixia and Yuan, Dongfeng},
  journal={IEEE Trans. Veh. Technol.},
  volume={72},
  number={10},
  pages={13173--13189},
  year={2023},
  publisher={IEEE}
}

@article{yan2023joint,
  title={Joint optimization of resource allocation and trajectory control for mobile group users in fixed-wing {UAV}-enabled wireless network},
  author={Yan, Xuezhen and Fang, Xuming and Deng, Cailian and Wang, Xianbin},
  journal={IEEE Trans. Wireless Commun.},
  volume={23},
  number={2},
  pages={1608--1621},
  year={2023},
  publisher={IEEE}
}

@article{omoniwa2022optimizing,
  title={Optimizing energy efficiency in {UAV}-assisted networks using deep reinforcement learning},
  author={Omoniwa, Babatunji and Galkin, Boris and Dusparic, Ivana},
  journal={IEEE Wireless Commun. Lett.},
  volume={11},
  number={8},
  pages={1590--1594},
  year={2022},
  publisher={IEEE}
}

@inproceedings{yang2021dynamic,
  title={Dynamic trajectory and offloading control of {UAV}-enabled {MEC} under user mobility},
  author={Yang, Zheyuan and Bi, Suzhi and Zhang, Ying-Jun Angela},
  booktitle={IEEE Int. Conf. Commun. Workshops (ICC Workshops)},
  pages={1--6},
  year={2021},
}

@article{liu2019trajectory,
  title={Trajectory design and power control for multi-{UAV} assisted wireless networks: A machine learning approach},
  author={Liu, Xiao and Liu, Yuanwei and Chen, Yue and Hanzo, Lajos},
  journal={IEEE Trans. Veh. Technol.},
  volume={68},
  number={8},
  pages={7957--7969},
  year={2019},
  publisher={IEEE}
}

@article{amer2020mobility,
  title={Mobility in the sky: Performance and mobility analysis for cellular-connected {UAV}s},
  author={Amer, Ramy and Saad, Walid and Marchetti, Nicola},
  journal={IEEE Trans. on Commun.},
  volume={68},
  number={5},
  pages={3229--3246},
  year={2020},
  publisher={IEEE}
}

@article{wu2024deep,
  title={Deep learning for secure {UAV} swarm communication under malicious attacks},
  author={Wu, Qirui and Zhang, Yirun and Yang, Zhaohui and Shikh-Bahaei, Mohammad R},
  journal={IEEE Trans. Wireless Commun.},
  volume={23},
  number={10},
  pages={14879--14894},
  year={2024},
  publisher={IEEE}
}

@article{bai2022delay,
  title={Delay-aware cooperative task offloading for multi-{UAV} enabled edge-cloud computing},
  author={Bai, Zhuoyi and Lin, Yifan and Cao, Yang and Wang, Wei},
  journal={IEEE Trans. Mob. Comput.},
  volume={23},
  number={2},
  pages={1034--1049},
  year={2022},
  publisher={IEEE}
}

@article{wang2021deep,
  title={Deep reinforcement learning based dynamic trajectory control for {UAV}-assisted mobile edge computing},
  author={Wang, Liang and Wang, Kezhi and Pan, Cunhua and Xu, Wei and Aslam, Nauman and Nallanathan, Arumugam},
  journal={IEEE Trans. Mob. Comput.},
  volume={21},
  number={10},
  pages={3536--3550},
  year={2021},
  publisher={IEEE}
}

@article{zeng2019energy,
  title={Energy minimization for wireless communication with rotary-wing {UAV}},
  author={Zeng, Yong and Xu, Jie and Zhang, Rui},
  journal={IEEE Trans. Wireless Commun.},
  volume={18},
  number={4},
  pages={2329--2345},
  year={2019},
  publisher={IEEE}
}

@article{yang2022online,
  title={Online trajectory and resource optimization for stochastic {UAV}-enabled {MEC} systems},
  author={Yang, Zheyuan and Bi, Suzhi and Zhang, Ying-Jun Angela},
  journal={IEEE Trans. Wireless Commun.},
  volume={21},
  number={7},
  pages={5629--5643},
  year={2022},
  publisher={IEEE}
}

@article{wang2025joint,
  title={Joint positioning and computation offloading in multi-{UAV} {MEC} for low latency applications: A proximal policy optimization approach},
  author={Wang, Yuhui and Farooq, Junaid and Ghazzai, Hakim},
  journal={IEEE Trans. Mob. Comput.},
  year={2025},
	volume={24},
  number={10},
  pages={9584--9598},
  publisher={IEEE}
}

@article{wang2021computation,
  title={Computation offloading optimization for {UAV}-assisted mobile edge computing: A deep deterministic policy gradient approach},
  author={Wang, Yunpeng and Fang, Weiwei and Ding, Yi and Xiong, Naixue},
  journal={Wirel. Netw.},
  volume={27},
  number={4},
  pages={2991--3006},
  year={2021},
  publisher={Springer}
}

@article{hu2019uav,
  title={{UAV}-assisted relaying and edge computing: Scheduling and trajectory optimization},
  author={Hu, Xiaoyan and Wong, Kai-Kit and Yang, Kun and Zheng, Zhongbin},
  journal={IEEE Trans. Wireless Commun.},
  volume={18},
  number={10},
  pages={4738--4752},
  year={2019},
  publisher={IEEE}
}

@article{hu2018joint,
  title={Joint offloading and trajectory design for {UAV}-enabled mobile edge computing systems},
  author={Hu, Qiyu and Cai, Yunlong and Yu, Guanding and Qin, Zhijin and Zhao, Minjian and Li, Geoffrey Ye},
  journal={IEEE Internet Things J.},
  volume={6},
  number={2},
  pages={1879--1892},
  year={2018},
  publisher={IEEE}
}

@article{ren2024resource,
  title={Resource allocation and slicing strategy for multiple services co-existence in wireless train communication network},
  author={Ren, Qiao and Lin, Siyu and Cai, Yifei and Deng, Xiaoheng and Kong, Linghe and Mumtaz, Shahid and Ai, Bo},
  journal={IEEE Trans. Wireless Commun.},
	volume={24},
  number={1},
  pages={401--414},
  year={2025},
  publisher={IEEE}
}

@article{lu2025agentic,
  title={Agentic graph neural networks for wireless communications and networking towards edge general intelligence: A survey},
  author={Lu, Yang and Zhang, Shengli and Liu, Chang and Zhang, Ruichen and Ai, Bo and Niyato, Dusit and Ni, Wei and Wang, Xianbin},
  journal={arXiv preprint arXiv:2508.08620},
  year={2025}
}

@article{tang2025deep,
  title={Deep graph reinforcement learning for {UAV}-enabled multi-user secure communications},
  author={Tang, Xiao and Zhao, Kexin and Shen, Chao and Du, Qinghe and Wang, Yichen and Niyato, Dusit and Han, Zhu},
  journal={IEEE Trans. Mob. Comput.},
	volume={24},
  number={9},
  pages={8780 - 8793},
  year={2025},
  publisher={IEEE}
}

@article{mao2024uav,
  title={{UAV}-assisted communications in {SAGIN-ISAC}: Mobile user tracking and robust beamforming},
  author={Mao, Weihao and Lu, Yang and Pan, Gaofeng and Ai, Bo},
  journal={IEEE J. Sel. Areas Commun.},
	volume={43},
  number={1},
  pages={186-200},
  year={2025},
  publisher={IEEE}
}

@article{sun2021joint,
  title={Joint computation offloading and trajectory planning for {UAV}-assisted edge computing},
  author={Sun, Chao and Ni, Wei and Wang, Xin},
  journal={IEEE Trans. Wireless Commun.},
  volume={20},
  number={8},
  pages={5343--5358},
  year={2021},
  publisher={IEEE}
}

@article{askari2021energy,
  title={Energy-efficient and real-time {NOMA} scheduling in {IoMT}-based three-tier {WBANs}},
  author={Askari, Zeinab and Abouei, Jamshid and Jaseemuddin, Muhammad and Anpalagan, Alagan},
  journal={IEEE Internet Things J.},
  volume={8},
  number={18},
  pages={13975--13990},
  year={2021},
  publisher={IEEE}
}

@article{paszke2019pytorch,
  title={Pytorch: An imperative style, high-performance deep learning library},
  author={Paszke, Adam and Gross, Sam and Massa, Francisco and Lerer, Adam and Bradbury, James and Chanan, Gregory and Killeen, Trevor and Lin, Zeming and Gimelshein, Natalia and Antiga, Luca and others},
  journal={Adv. Neural Inf. Process. Syst.},
  volume={32},
  year={2019}
}

@article{xu2017online,
  title={Online learning for offloading and autoscaling in energy harvesting mobile edge computing},
  author={Xu, Jie and Chen, Lixing and Ren, Shaolei},
  journal={IEEE Trans. Cogn. Commun. Netw.},
  volume={3},
  number={3},
  pages={361--373},
  year={2017},
  publisher={IEEE}
}

@article{chen2025computation,
  title={Computation offloading optimization for {UAV}-based cloud-edge collaborative task scheduling strategy},
  author={Chen, Haosheng and Cui, Haixia and Wang, Jiahuan and Cao, Peng and He, Yejun and Guizani, Mohsen},
  journal={IEEE Trans. Cogn. Commun. Netw.},
  year={2025},
  publisher={IEEE}
}

@article{liu2023energy,
  title={Energy efficient computation offloading in aerial edge networks with multi-agent cooperation},
  author={Liu, Wenshuai and Li, Bin and Xie, Wancheng and Dai, Yueyue and Fei, Zesong},
  journal={IEEE Trans. Wireless Commun.},
  volume={22},
  number={9},
  pages={5725--5739},
  year={2023},
  publisher={IEEE}
}

@article{philip2021internet,
  title={{Internet} of {Things} for in-home health monitoring systems: Current advances, challenges and future directions},
  author={Philip, Nada Y and Rodrigues, Joel JPC and Wang, Honggang and Fong, Simon James and Chen, Jia},
  journal={IEEE J. Sel. Areas Commun.},
  volume={39},
  number={2},
  pages={300--310},
  year={2021},
  publisher={IEEE}
}

@article{movassaghi2014wireless,
  title={Wireless body area networks: A survey},
  author={Movassaghi, Samaneh and Abolhasan, Mehran and Lipman, Justin and Smith, David and Jamalipour, Abbas},
  journal={IEEE Commun. Surv. Tutor.},
  volume={16},
  number={3},
  pages={1658--1686},
  year={2014},
  publisher={IEEE}
}

\end{document}